\documentclass[12pt,preprint]{emulateapj}

\newcommand{\be}{\begin{equation}}
\newcommand{\ee}{\end{equation}}
\newcommand{\ba}{\begin{eqnarray}}
\newcommand{\ea}{\end{eqnarray}}

\begin{document}
\title{foreground contamination in Ly$\alpha$ intensity mapping during the epoch of reionization}

\author{Yan Gong$^1$, Marta Silva$^{2,3}$, Asantha Cooray$^1$, and Mario G. Santos$^{2,3,4}$}

\affil{$^1$Department of Physics \& Astronomy, University of California, Irvine, CA 92697}
\affil{$^2$CENTRA, Instituto Superior Tecnico, Technical University of Lisbon, Lisboa 1049-001, Portugal}
\affil{$^3$Department of Physics, University of Western Cape, Cape Town 7535, South Africa}
\affil{$^4$SKA SA, 3rd Floor, The Park, Park Road, Pinelands, 7405, South Africa}

\begin{abstract}

The intensity mapping of Ly$\alpha$ emission during the epoch of reionization (EoR) will be contaminated by foreground emission lines from lower redshifts. We calculate the mean intensity and the power spectrum of Ly$\alpha$ emission at $z\sim7$, and estimate the uncertainties according to the relevant astrophysical processes. We find that the low-redshift emission lines from $6563\ \rm \AA$ H$\alpha$, $5007\ \rm \AA$ [OIII] and $3727\ \rm \AA$ [OII] will be strong contaminants on the observed Ly$\alpha$ power spectrum. We make use of both the star formation rate (SFR) and luminosity functions (LF) to estimate the mean intensity and power spectra of the three foreground lines at $z\sim0.5$ for H${\alpha}$, $z\sim0.9$ for [OIII] and $z\sim1.6$ for [OII], as they will contaminate the Ly$\alpha$ emission at $z\sim7$. The [OII] line is found to be the strongest. We analyze the masking of the bright survey pixels with a foreground line above some line intensity threshold as a way to reduce the contamination in an intensity mapping survey. We find that the foreground contamination can be neglected if we remove pixels with fluxes above $1.4\times 10^{-20}\ \rm W/m^2$.

\end{abstract}


\maketitle

\section{Introduction}

The epoch of reionization (EoR) is an important and largely unconstrained stage in the evolution of our universe \citep{Barkana01}. The first stars and galaxies can start ionization of the neutral intergalactic medium (IGM) as early as $z\sim 20$ or even before while completion of the ionization process should end much later, at a redshift as low as $z\sim6$. This is suggested by the observations of the absorption of the Ly$\alpha$ emission from quasars by the neutral hydrogen in the IGM \citep{Fan06}, i.e. the Gunn-Peterson effect \citep{Gunn65}. Most of the fundamental stones of this era are still poorly known, such as the history, the formation and evolution of the stars and galaxies, and the formation of large scale structure (LSS).

One method to study the EoR is through 21-cm emission of neutral hydrogen in the IGM \citep{Furlanetto06}, which provides a direct method to study the ionization history of the IGM. However, the 21-cm measurement can not directly trace the galaxy distribution or reflect the star formation. It also suffers from foreground contamination at several orders of magnitude higher than the cosmological 21-cm signal. These disadvantages restrict the 21-cm measurements to explore the LSS and the processes of the formation and evolution for stars and galaxies during the EoR. 

A complementary measurement to 21-cm observations is the intensity mapping of the emission lines. This approach is more closely connected to the galaxy properties. The emission lines generated from the stars and molecular or atomic gas in galaxies are more sensible probes of the metallicity and star formation of galaxies. They trace the galaxy distribution at large scales, providing information about the LSS. Several emission lines from galaxies have been proposed to study the EoR \citep{Visbal10}, e.g. CO rotational lines \citep{Righi08,Gong11,Carilli11,Lidz11}, [CII] fine structure line \citep{Gong12}, and the Ly$\alpha$ line \citep{Silva13,Pullen13}. Even molecular hydrogen H$_2$ in the pre-reionization era can be used to study the formation of first stars and galaxies \citep{Gong13}.

The intensity mapping technique is a powerful tool to explore large areas at poor spatial or angular resolution. It does not attempt to resolve individual sources, but measures the cumulative emission from all sources. Thus, intensity mapping of emission lines from galaxies provides a suitable way to study the statistical properties of galaxies during the EoR in an acceptable survey timescale. However, intensity maps are easily contaminated by other emission lines from lower redshifts in the same observed frequency range. Since intensity mapping cannot resolve individual sources, it is important to independently identify and eliminate the contamination from the foreground lines.

In this work, we study the foreground contamination on Ly$\alpha$ intensity maps of the EoR. We first compute the Ly$\alpha$ emission from both galaxies and the IGM due to recombination and collision processes. We calculate the Ly$\alpha$ mean intensity and power spectrum at $z\sim7$. We find the $6563\ \rm\AA$ H$\alpha$, $5007\ \rm\AA$ [OIII] and $3727\ \rm \AA$ [OII] are the strongest contamination on the Ly$\alpha$ observations. We estimate the mean intensities of these three lines using the star formation rate (SFR) derived from observations and simulations. Also, for comparison, we make use of luminosity functions (LFs) from different observations to calculate the mean intensities and derive their anisotropy power spectra. We then discuss ways to remove contamination by masking the bright pixels of the intensity maps.

The paper is organized as follows: in the next Section, we estimate the Ly$\alpha$ emission from both galaxies and the IGM, and calculate the mean intensity and power spectrum during the EoR. In Section 3, we explore the intensity and power spectra of the foreground emission lines of the H$\alpha$, [OII] and [OIII] around $z=0.5$, $0.9$ and $1.6$ that can contaminate the Ly$\alpha$ emission at $z\sim7$, and derive their mean intensity and power spectra. In Section 4, we discuss the projection effect of the foreground power spectra and investigate the methods of removing the foreground contamination from low-redshift emission lines. We summarize our results in Section 5. We assume the flat $\Lambda$CDM with $\Omega_{\rm M}=0.27$, $\Omega_{\rm b}=0.046$, and $h=0.71$ for the calculation throughout the paper \citep{Komatsu11}.

\section{The Ly$\alpha$ intensity and anisotropy power spectrum during the EoR}

In this Section, we estimate the mean intensity and power spectrum of the Ly$\alpha$ emission during the EoR. The processes that originate Ly$\alpha$ in galaxies and in the IGM are mainly recombinations and collisions. The other processes, e.g. gas cooling by the falling of the IGM gas into the potential wells of the dark matter, and continuum emission by stellar, free-free, free-bound and two-photon emission, can be safely neglected according to previous studies \citep{Cooray12,Silva13}. 

\subsection{The Ly$\alpha$ mean intensity}

Following \cite{Silva13}, we estimate the luminosity of the Ly$\alpha$ emission from galaxies by the recombination and collision processes as
\ba\label{eq:L_rec}
L_{\rm rec}^{\rm gal}(M,z) = &1.55&\times 10^{42}\ f_{\rm Ly\alpha}(z)\ [1-f_{\rm esc}^{\rm ion}(M,z)] \nonumber \\
                      &\times& \frac{{\rm SFR}(M,z)}{M_{\sun}\ \rm yr^{-1}}\ ({\rm erg\ s^{-1}}),
\ea
\ba\label{eq:L_coll}
L_{\rm coll}^{\rm gal}(M,z) = &4.03&\times 10^{41}\ f_{\rm Ly\alpha}(z)\ [1-f_{\rm esc}^{\rm ion}(M,z)] \nonumber \\
                      &\times& \frac{{\rm SFR}(M,z)}{M_{\sun}\ \rm yr^{-1}}\ ({\rm erg\ s^{-1}}),
\ea
respectively \footnote{Note that the ratio $L_{\rm coll}^{\rm gal}/L_{\rm rec}^{\rm gal}=0.26$ can be larger at high redshifts due to strong galaxy forming process \citep{Laursen13}. This ratio is still observationally unconstrained, and also has large uncertainty suggesting by simulations.}. Here the $f_{\rm Ly\alpha}=10^{-3}\times C_{\rm dust}(1+z)^{\xi}$ is the fraction of the Ly$\alpha$ photons which are not absorbed by the dust in the galaxy, where $C_{\rm dust}=3.34$ and $\xi=2.57$ \citep{Hayes11}. Note that this fitting formula is mainly derived from low-redshift sources and extrapolated to high redshifts ($z>6$). The $f_{\rm Ly\alpha}$ can be much lower at $z\sim7$ \citep[e.g.][]{Stark10, Stark11}. The $f_{\rm esc}^{\rm ion}={\rm exp}[-\alpha(z)M^{\beta(z)}]$ is the escape fraction of the ionizing photons, where $M$ is the halo mass, and $\alpha=5.18\times 10^{-3}$ and $\beta=0.244$ \citep{Razoumov10,Silva13}. The ${\rm SFR}(M,z)$ is the star formation rate, which is parameterized as
\ba
{\rm SFR}(M,z=7) &=& 1.6\times 10^{-26}\ M^a\left(1+\frac{M}{c_1}\right)^b \nonumber \\
            &\times& \left(1+\frac{M}{c_2}\right)^d\left(1+\frac{M}{c_3}\right)^e,
\ea
where $a=2.59$, $b=-0.62$, $c_1=8\times10^8\ M_{\sun}$, $c_2=7\times 10^{9}\ M_{\sun}$, $c_3=10^{11}\ M_{\sun}$, $d=0.4$ and $e=-2.25$. This SFR parameterization was derived to fit the properties of the simulated galaxies catalogs from De Lucia et al. (2007) and Guo et al. (2011),and is available at $z=7$ in \citep{Silva13}. Then the integrate Ly$\alpha$ mean intensity from individual galaxies is given by
\be \label{eq:I_gal}
\bar{I}_{\rm gal}(z) = \int_{M_{\rm min}}^{M_{\rm max}} dM\frac{dn}{dM}\frac{L_{\rm gal}(M,z)}{4\pi D_{\rm L}^2} y(z)D_{\rm A}^2,
\ee
where $L^{\rm gal}=L_{\rm rec}^{\rm gal}+L_{\rm coll}^{\rm gal}$ is the total luminosity of Ly$\alpha$ emission from galaxies, $dn/dM$ is the halo mass function \citep{Sheth99}, and $D_{\rm L}$ and $D_{\rm A}$ are the luminosity and comoving angular diameter distance respectively. The factor $y(z)=dr/d\nu=\lambda_{{\rm Ly}\alpha}(1+z)^2/H(z)$, where $r(z)$ is the comoving distance at $z$, $\lambda_{{\rm Ly}\alpha}=1216\ \rm \AA$ is the Ly$\alpha$ wavelength in the rest frame, and $H(z)$ is the Hubble parameter. We take $M_{\rm min}=10^8\ h^{-1}M_{\sun}$ and $M_{\rm min}=10^{13}\ h^{-1}M_{\sun}$, which denote the mass range of the halos that host the galaxies with the Ly$\alpha$ emission in this work.

For the Ly$\alpha$ emission from the IGM, we use the luminosity density to estimate the mean intensity
\be \label{eq:I_IGM}
\bar{I}_{\rm IGM} = \frac{l^{\rm IGM}(z)}{4\pi D_{\rm L}^2}\ y(z)D_{\rm A}^2,
\ee
where $l^{\rm IGM}=l^{\rm IGM}_{\rm rec}+l^{\rm IGM}_{\rm coll}$ is the Ly$\alpha$ luminosity density for the IGM in $\rm erg\ s^{-1}cm^{-3}$. The $l^{\rm IGM}_{\rm rec}$ and $l^{\rm IGM}_{\rm coll}$ are the luminosity densities from recombination and collision processes, respectively. The $l^{\rm IGM}_{\rm rec}$ can be estimated by $l^{\rm IGM}_{\rm rec} = \epsilon_{\rm Ly\alpha}^{\rm rec}h\nu_{\rm Ly\alpha}$, where $\epsilon_{\rm Ly\alpha}^{\rm rec}$ is the Ly$\alpha$ recombination emission rate per cm$^3$, which is given by
\be
\epsilon_{\rm Ly\alpha}^{\rm rec}=f_{\rm Ly\alpha}^{\rm rec}n_en_{\rm HII}\alpha_{\rm B}^{\rm rec}.
\ee
The $n_e$ and $n_{\rm HII}$ are the number density of the electron and HII. Here we have $n_e n_{\rm HII}=C_{\rm IGM}(z)\bar{n}_e(z) \bar{n}_{\rm HII}(z)$, where $\bar{n}_e(z)$ and $\bar{n}_{\rm HII}(z)$ are the mean number density of the electron and HII at z, which are dependent on the reionization fraction $x_{\rm i}$ during the EoR. The $C_{\rm IGM}(z)=\langle n^2\rangle/\langle n\rangle^2$ is the clumping factor of the IGM at z which can be set by $C_{\rm IGM}=6$ at z=7 \citep{Pawlik09,Silva13}. The $\alpha_{\rm B}^{\rm rec}$ is the hydrogen case B recombination coefficient which can be fitted by \citep{Hummer94,Seager99}
\be
\alpha_{\rm B}^{\rm rec}(T)=10^{-13}\frac{aT_4^b}{1+cT_4^d}\ \rm (cm^3/s),
\ee
where $a=4.309$, $b=-0.6166$, $c=0.6703$, $d=0.5300$, and $T_4=T/10^4$ K. The $f_{\rm Ly\alpha}^{\rm rec}$ is the the fraction of the Ly$\alpha$ photons produced in the case B recombination, and we use the fitting formula from \cite{Cantalupo08} to evaluate it as
\be
f_{\rm Ly\alpha}^{\rm rec}(T) = 0.686-0.106{\rm log_{10}}(T_4)-0.009T_4^{-0.44}.
\ee
This formula is accurate to 0.1\% for $100<T<10^5{\rm K}$. We assume the mean gas temperature of the IGM to be $1.5\times 10^4$ K in this work, which is in a good agreement with the results from simulations and the current measurements from quasars \citep{Theuns02,Tittley07,Trac08,Bolton10,Bolton12}.

The Ly$\alpha$ collisional emission in the IGM involves free electrons that collide with the neutral hydrogen atoms (HI) and transfer their kinetic energy by exciting the HI to high energy levels. The hydrogen atoms then decay by emitting photons, including Ly$\alpha$ photons. During the EoR, this process mainly occurs in the ionizing fronts of the reionization bubbles, since this emission will be the strongest when $n_e\sim n_{\rm HI}$. Similar to the recombination emission, the luminosity density of the collisional emission in the IGM can be calculated by $l^{\rm IGM}_{\rm coll}=\epsilon_{\rm Ly\alpha}^{\rm coll}h\nu_{\rm Ly\alpha}$. The $\epsilon_{\rm Ly\alpha}^{\rm coll}$ is the collisional emission rate per cm$^3$
\be \label{eq:coll_emi_IGM}
\epsilon_{\rm Ly\alpha}^{\rm coll} = C_{\rm Ly\alpha}^{\rm eff}n_en_{\rm HI},
\ee
where $n_{\rm HI}$ is the number density of the neutral hydrogen atoms. We assume $n_en_{\rm HI}=f^{\rm ion}_{\rm front}C_{\rm IGM}(z)\bar{n}_e(z) \bar{n}_{\rm HI}(z)$, where $\bar{n}_{\rm HI}(z)$ is the mean number density of neutral hydrogen at z. The $f^{\rm ion}_{\rm front}$ is the volume fraction of the bubble ionizing fronts. Because the ionizing front just takes a very small part of the whole ionizing bubble, as a good approximation, the volume fraction of the ionizing front can be estimated as $f^{\rm ion}_{\rm front}=\frac{4}{3}\pi(r_2^3-r_1^3)$, where $r_1=[(3/4\pi)x_{\rm i}]^{1/3}$, and $r_2=r_1+d$ and $d=\frac{1}{6}[n_{\rm H}\sigma_{\rm H}(\langle\nu \rangle)]^{-1}$ is the thickness of the bubble ionizing front. The factor $1/6$ is derived from the integration of $x_{\rm HI}(1-x_{\rm HI})$ over the ionizing front where $x_{\rm HI}(r)$ is the fraction of the HI in the front at radius $r$ \citep{Cantalupo08}. Here $\sigma_{\rm H}$ denotes the cross section of the hydrogen ionization, where $\langle\nu\rangle$ is the mean frequency of the ionizing photons, and it is given by \citep{Osterbrock89}
\be
\sigma_{\rm H}(\nu) = \frac{\sigma_0}{1-{\rm exp}(-2\pi/\epsilon)}\left[\frac{\nu_0}{\nu}{\rm exp}\left( 1-\frac{{\rm tan^{-1}}\epsilon}{\epsilon}\right) \right]^4.
\ee
Here $\nu_0$ is the Lyman limit frequency, $\sigma_0=6.3\times 10^{-18}\ \rm cm^2$ and $\epsilon\equiv (\nu/\nu_0-1)^{1/2}$. Here we take $\langle\nu \rangle=2.4\nu_0$ in our calculation \citep{Gould96}. Note that we set $f^{\rm ion}_{\rm front}=0$ when $x_{\rm i}\ge1$, which means there is no collisional emission in the IGM when the universe is totally ionized. 

In Eq.~(\ref{eq:coll_emi_IGM}), the $C_{\rm Ly\alpha}^{\rm eff}$ is the effective collisional excitation coefficient, which can be estimated by \citep{Cantalupo08}
\be
C_{\rm Ly\alpha}^{\rm eff} = C_{1,2p}+C_{1,3s}+C_{1,3d} \, .
\ee
Here we take into account the excitation up to $n=3$ energy level to decay and produce Ly$\alpha$ photons. The contribution of emission from higher energy levels can be neglected at $z\approx7$ given that the gas temperature and density of the IGM are on average too low to excite these levels. The collisional excitation rate $C_{\rm l,u}$, in cm$^3$ per second, is given by
\be
C_{\rm l,u} = \frac{8.629\times 10^{-6}}{g_{\rm l}\sqrt{T}}\gamma_{\rm l,u}(T){\rm exp}\left( \frac{-\Delta E_{\rm l,u}}{kT}\right)\ \rm (cm^3/s),
\ee
where $T$ is the gas temperature that we assume $T=1.5\times 10^4$ K, $\Delta E_{\rm l,u}$ is the energy difference between lower level and higher level, $g_{\rm l}$ is the statistic weight for lower level, and $\gamma_{\rm l,u}(T)$ is the effective collision strength calculated by the fitting formulae from \cite{Giovanardi87}.

\begin{figure}[htpb]
\includegraphics[scale = 0.4]{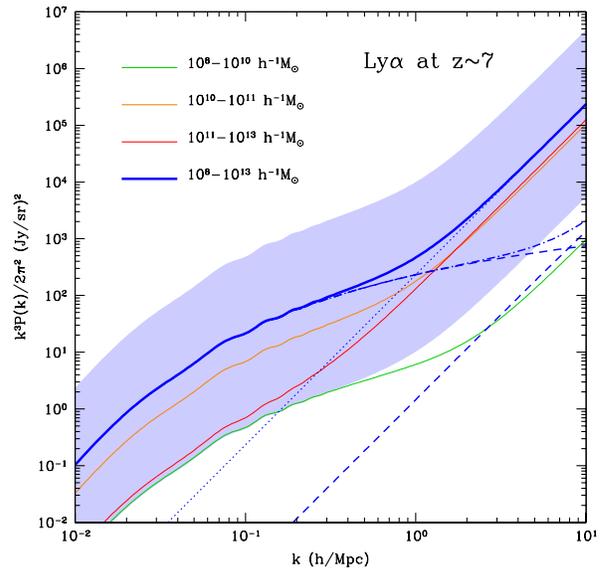}
\caption{\label{fig:P_Lya} The power spectrum of the Ly$\alpha$ emission at $z=7$. The dash-dotted line is the clustering power spectrum which consists of 1-halo and 2-halo components in dashed lines, the dotted line denotes the shot-noise power spectrum, and the total power spectrum is shown in solid line. The uncertainty of the power spectrum is shown in shaded region which is estimated by the uncertainties of the $f_{\rm esc}^{\rm ion}$, $f_{\rm Ly\alpha}$, SFR and IGM clumping factor. The contributions of different halo mass scales to the total power spectrum are also shown. We find $10^{10}$-$10^{11}\ h^{-1}M_{\sun}$ dominates the clustering power spectrum.}
\end{figure}

Then we estimate the total mean intensity of the Ly$\alpha$ emission $\bar{I}_{\rm Ly\alpha}=\bar{I}_{\rm gal}+\bar{I}_{\rm IGM}$ using Eq.~(\ref{eq:I_gal}) and Eq.~(\ref{eq:I_IGM}). At $z\sim 7$, we get $\bar{I}_{\rm gal}=9.2\ \rm Jy/sr$ and $\bar{I}_{\rm IGM}=1.2\ \rm Jy/sr$ if assuming {\bf $\bar{x}_{\rm i}=0.85$}. Thus, according to the assumptions we have made, Ly$\alpha$ emission from galaxies is larger than that from the IGM around $z=7$. Note that the collisional emission in the IGM is much smaller than the recombination emission with $\bar{I}^{\rm IGM}_{\rm coll}\simeq 0.1\ \rm Jy/sr$ and  $\bar{I}^{\rm IGM}_{\rm rec}\simeq 1.1\ \rm Jy/sr$,respectively, assuming $T=1.5\times 10^4$ K for the gas temperature of the IGM. 


There are large uncertainties for the parameters in the estimation of the Ly$\alpha$ intensity. According to the current measurements \citep{Razoumov10, Hayes11, Blanc11}, we find the $f_{\rm esc}^{\rm ion}$ and $f_{\rm Ly\alpha}$ \footnote{This $f_{\rm Ly\alpha}$ is Ly$\alpha$ ``effective'' escape fraction instead of real escape fraction which should include the scattered Ly$\alpha$ emission from the diffuse halos surrounding galaxies \citep{Dijkstra13}. The Ly$\alpha$ emission from the diffuse halos has low surface brightness, but the total flux can exceed the direct emission from galaxies by a significant factor \citep[e.g.][]{Zheng10,Steidel10,Matsuda12}. Thus the real Ly$\alpha$ escape fraction can be much larger than the effective escape fraction. This diffused emission could contribute to the intensity mapping. We find our fiducial $f_{\rm Ly\alpha}\simeq0.7$ at $z=7$ which is close to one, and we also add an uncertainty of a factor of 2 on it. So the effect of the diffused Ly$\alpha$ emission is safely covered in our calculation.} in Eq.~(\ref{eq:L_rec}) and (\ref{eq:L_coll}) have an uncertainty of a factor of 2 each. Besides, we take into account of the uncertainties in the SFR and the clumping fator, which can be factors of 3 and 5 at $z\sim7$ respectively \citep{Silva13}. Eventually, we find $\bar{I}_{\rm Ly\alpha}=10.4^{+39.3}_{-9.0}\ \rm Jy/sr$ around $z=7$. Our results are well consistent with other works when allowing for uncertainty (e.g. Silva et al. 2013; Pullen et al. 2013). Besides, we note that the estimated Ly$\alpha$ emission of high-$z$ metal-poor galaxies with standard case-B assumption can be underestimated by a factor of $\sim 2-3$ compared to accurate prediction \citep{Raiter10}. This possible departure can be covered by the intensity uncertainty we consider.

\subsection{The Ly$\alpha$ power spectrum}

According to the calculations above, the absolute value of the Ly$\alpha$ intensity background during the EoR is small and hard to measure directly in an absolute intensity experiment. Instead, we can try to measure the fluctuations of the Ly$\alpha$ intensity and estimate the anisotropy power spectrum. Since the Ly$\alpha$ emission from galaxies and IGM in the ionization bubbles surrounding the galaxies trace the underlying matter density field, we can calculate the Ly$\alpha$ intensity fluctuations by
\be
\delta I_{\rm Ly\alpha} = \bar{b}_{\rm Ly\alpha} \bar{I}_{\rm Ly\alpha}\delta(\rm \bf x).
\ee
Here we set $\bar{I}_{\rm Ly\alpha}\simeq \bar{I}_{\rm gal}$, since the main Ly$\alpha$ emission at $z\sim 7$ comes from galaxies. The $\delta(\rm \bf x)$ is the matter over-density at the position $\rm \bf x$, and $\bar{b}_{\rm Ly\alpha}(z)$ is the average galaxy bias weighted by the Ly$\alpha$ luminosity 
\be \label{eq:b_Lya}
\bar{b}_{\rm Ly\alpha}(z)=\frac{\int^{M_{\rm max}}_{M_{\rm min}} dM \frac{dn}{dM} L_{\rm gal}^{\rm Ly\alpha} b(M,z)}{\int^{M_{\rm max}}_{M_{\rm min}} dM \frac{dn}{dM} L_{\rm gal}^{\rm Ly\alpha}},
\ee
where $L_{\rm gal}^{\rm Ly\alpha}(M,z)$ is the Ly$\alpha$ luminosity of the galaxy, and $b(M,z)$ is the halo bias \citep{Sheth99}. Then we can calculate the Ly$\alpha$ clustering power spectrum due to galaxy clustering as
\be\label{eq:P_Lya}
P_{\rm Ly\alpha}^{\rm clus}(k,z) = \bar{b}_{\rm Ly\alpha}^2 \bar{I}_{\rm Ly\alpha}^2 P_{\delta \delta}(k,z),
\ee
where $P_{\delta \delta}(k,z)$ is the matter power spectrum which can be estimated from the halo model \citep{Cooray02}. The clustering power spectrum dominates the fluctuations at large scales. At small scales the Poisson noise caused by the discrete distribution of galaxies becomes important. This Poisson or shot-noise power spectrum takes the form (e.g. Visbal \& Loeb 2010; Gong et al. 2011):
\be\label{eq:Pshot_Lya}
P^{\rm shot}_{\rm Ly\alpha}(z) = \int_{M_{\rm min}}^{M_{\rm max}} dM \frac{dn}{dM} \left[\frac{L_{\rm gal}^{\rm Ly\alpha}}{4\pi D_{\rm L}^2}y(z)D_{\rm A}^2\right]^2.
\ee
Therefore, the total power spectrum can be written by $P_{\rm Ly\alpha}^{\rm tot}(k,z)=P_{\rm Ly\alpha}^{\rm clus}(k,z)+P^{\rm shot}_{\rm Ly\alpha}(z)$. In Figure~\ref{fig:P_Lya}, we show the Ly$\alpha$ power spectrum at $z=7$. We show the total, clustering and shot-noise power spectrum in solid, dash-dotted and dotted line, respectively. The dashed lines denote the 1-halo and 2-halo terms of the clustering power spectrum. The uncertainty of the total power spectrum is also shown in shaded region, which is derived from the uncertainties of the $f_{\rm esc}^{\rm ion}$, $f_{\rm Ly\alpha}$, SFR and IGM clumping factor. We also show the contributions of different halo mass scales to the total power spectrum in colored solid lines. As can be seen, the halos with mass $10^{10}$-$10^{11}\ h^{-1}M_{\sun}$ provide the most contribution on the clustering power spectrum, and the halos with higher masses ($10^{11}$-$10^{13}\ h^{-1}M_{\sun}$) have large shot-noise since they are bright and rare. We find $\bar{b}_{\rm Ly\alpha}=5.0$ and $P^{\rm shot}_{\rm Ly\alpha}=4.6\times 10^3\ \rm (Jy/sr)^2 (Mpc/h)^{-3}$ at $z=7$, and the shot-noise power spectrum dominates the total power spectrum at $k\gtrsim1\ {\rm Mpc^{-1}}h$.

\section{The estimation of the foreground emission lines}

Since we can not resolve individual sources with intensity mapping, the measurements of the Ly$\alpha$ emission during the EoR can be contaminated by emission lines from lower redshifts. Here we consider three low-redshift emission lines, H$\alpha$ at 6563 $\rm \AA$, [OIII] at 5007 $\rm \AA$ and [OII] at 3727 $\rm \AA$. At $z\sim 7$, the frequency of the Ly$\alpha$ line is about 300 THz, which can be then contaminated by H$\alpha$ at $z\sim 0.5$, [OIII] at $z\sim 0.9$ and [OII] at $z\sim 1.6$, respectively. We will estimate the mean intensity and power spectra of these lines in this Section.

\subsection{The mean intensity from the SFR}

The H$\alpha$ line at 6563 $\rm \AA$, [OIII] line at 5007 $\rm \AA$ and [OII] line at 3727 $\rm \AA$ are good tracers of the SFR of galaxies. The luminosity of these lines is related to the SFR as 
\ba
{\rm SFR}\,(M_{\sun}{\rm yr^{-1}}) &=& (7.9\pm2.4)\times 10^{-42} L_{\rm H\alpha},\label{eq:SFR_Ha}\\
{\rm SFR}\,(M_{\sun}{\rm yr^{-1}}) &=& (1.4\pm 0.4)\times 10^{-41} L_{\rm [OII]},\label{eq:SFR_OII}\\
{\rm SFR}\,(M_{\sun}{\rm yr^{-1}}) &=& (7.6\pm3.7)\times 10^{-42} L_{\rm [OIII]}.\label{eq:SFR_OIII}
\ea
The luminosity here is in $\rm erg\ s^{-1}$, and the relations for H$\alpha$ and [OII] are from \cite{Kennicutt98}, and the [OIII] relation is given in \cite{Ly07}. The SFR-$L_{\rm H\alpha}$ relation assumes the initial mass function from \cite{Salpeter55}, and we add $30\%$ uncertainty to it. These conversions are also in good agreement with other works\footnote{The SFR-L$\rm_{[OII]}$ relation has a larger uncertainty, and we will discuss it in Section 3.2.} (e.g. Hopkins et al. 2003; Wijesinghe et al. 2011; Drake et al. 2013). With the help of the relation of the SFR and the halo mass as a function of redshift SFR$(M,z)$, we can use these conversions to derive $L(M,z)$ and compute the mean intensities for these lines using Eq.~(\ref{eq:I_gal}).

\begin{figure}[htpb]
\includegraphics[scale = 0.4]{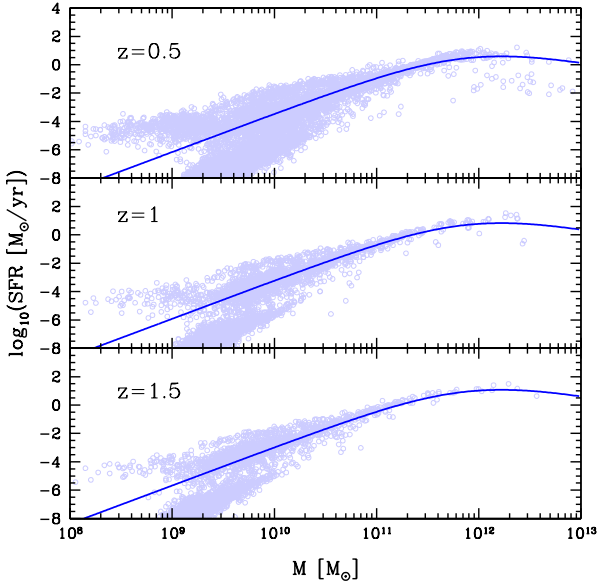}
\caption{\label{fig:SFR} The SFR vs. halo mass $M$ at $z=0.5$, $z=1$ and $z=1.5$ from the simulations. The dots are from the galaxy catalog in the simulations of \cite{Guo11} and the curves are the best fits of the SFR(M).}
\end{figure}

\begin{figure*}[htpb]
\epsscale{1.9}
\centerline{
\resizebox{!}{!}{\includegraphics[scale=0.29]{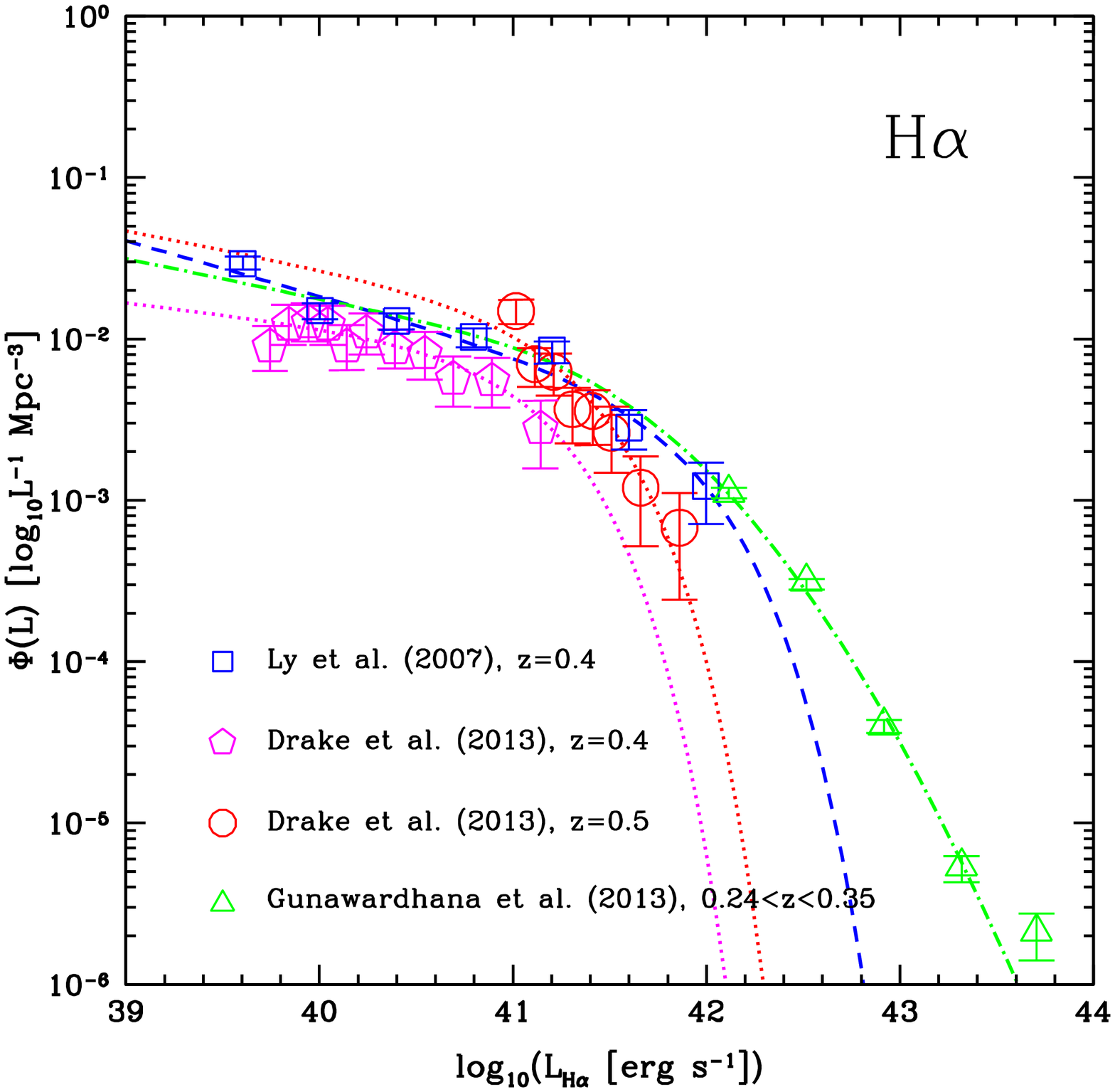}}
\resizebox{!}{!}{\includegraphics[scale=0.29]{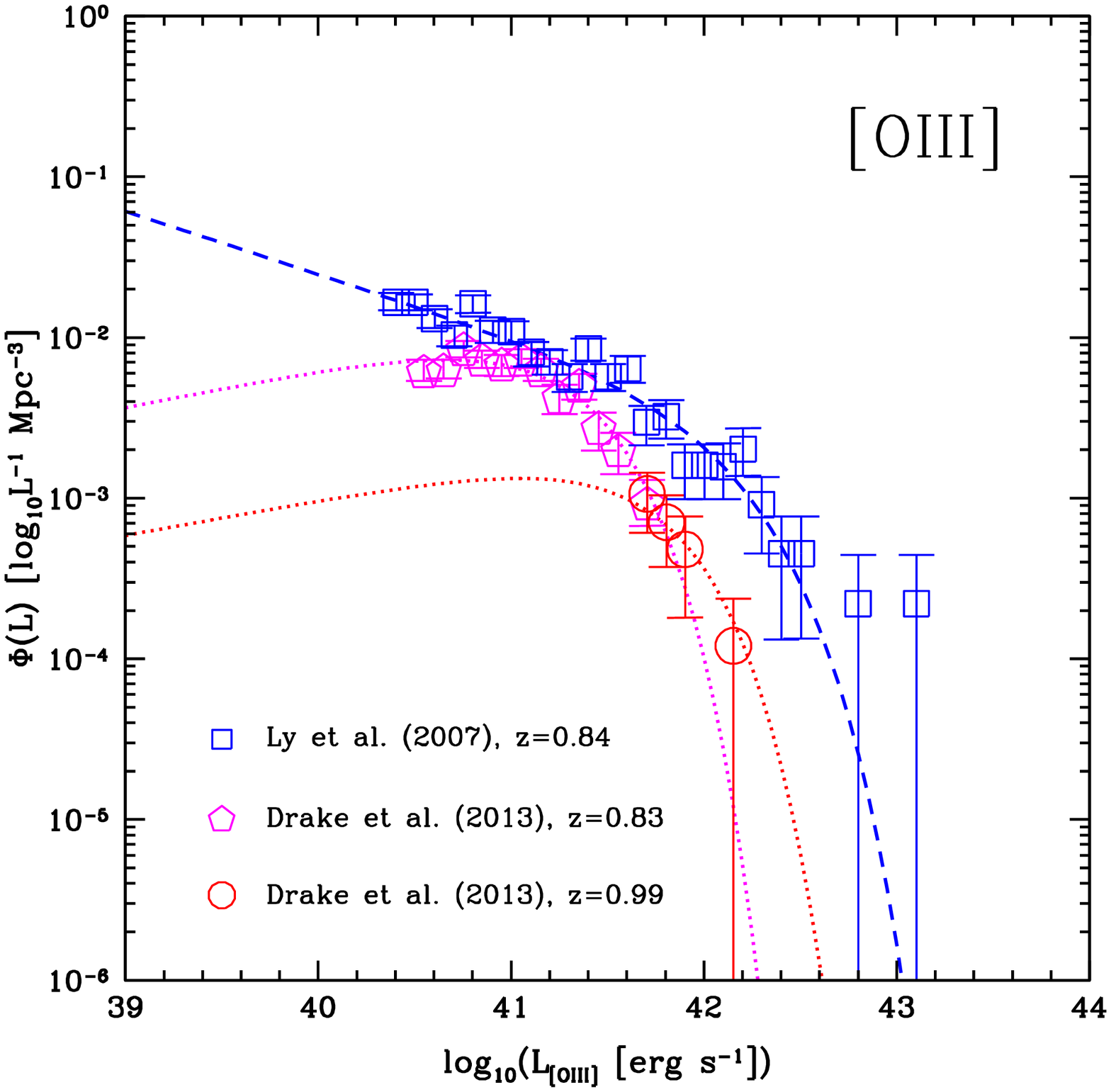}}
\resizebox{!}{!}{\includegraphics[scale=0.29]{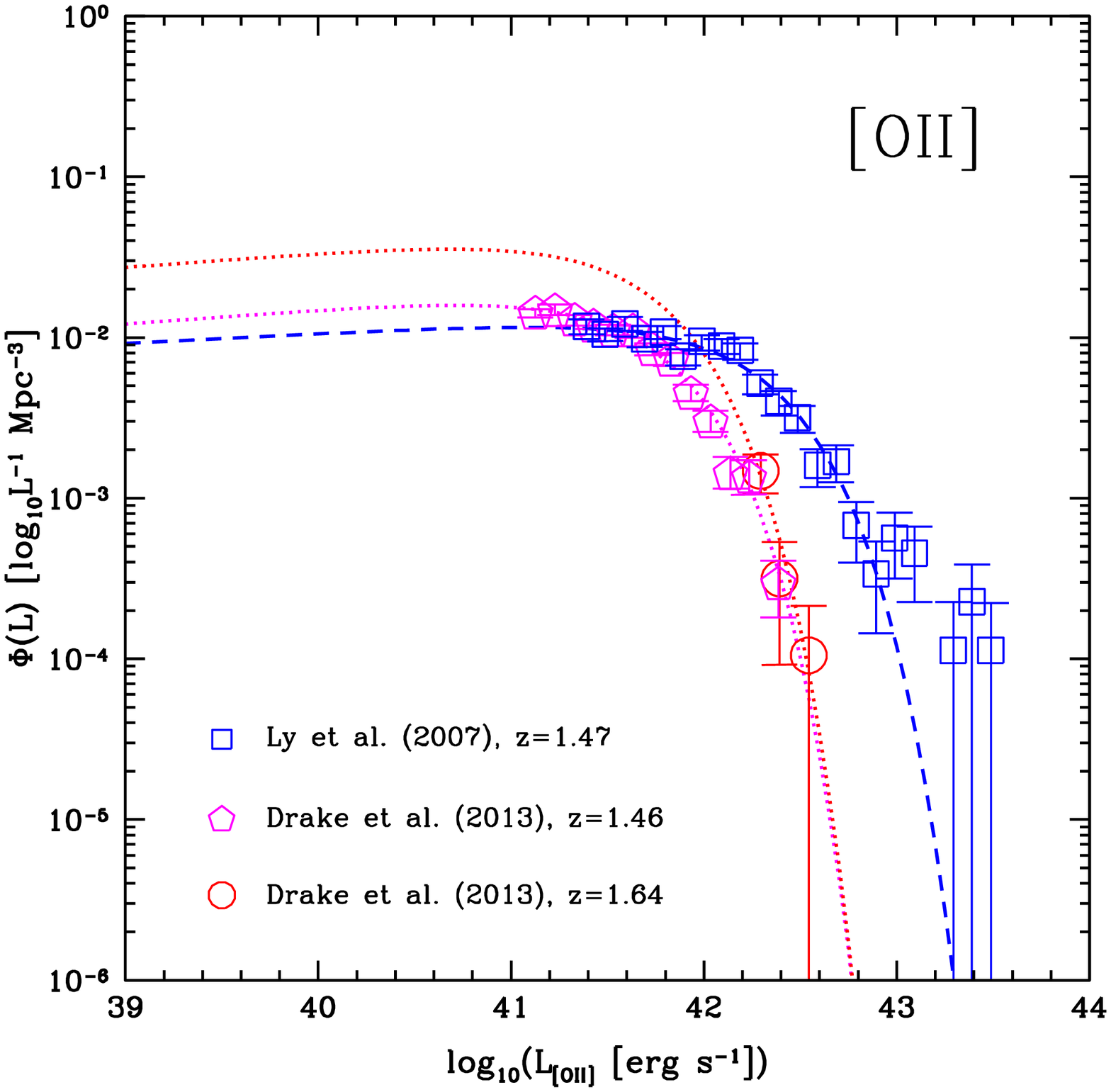}}
}
\epsscale{1.0}
\caption{\label{fig:LF} The luminosity functions and the best fits for the H$\alpha$, [OIII] and [OII] lines around $z=0.5$, $z=0.9$ and $z=1.6$ that could contaminate the Ly$\alpha$ emission at $z\sim7$. The blue squares, pink pentagons, red circles and green triangles are the observational LFs from Ly et al. (2007), Drake et al. (2013) and Gunawardhana et al. (2013), respectively. The dashed and dotted curves are the best fits of the corresponding LFs. Note that the data points of LFs from Drake et al. (2013) are the observed LFs (without dust extinction correction) which seem lower than the results of the other two works.}
\end{figure*}

To estimate the SFR$(M,z)$, we can make use of the SFRD$(z)$ from the observations and assume the SFR is proportional to the halo mass $M$. We take the SFRD$(z)$ given by \cite{Hopkins06} with the fitting formula from \cite{Cole01}
\be
{\rm SFRD}(z) = \frac{a+bz}{1+(z/c)^d}\ (h\,M_{\sun}\rm yr^{-1} Mpc^{-3}),
\ee
where $a=0.0118$, $b=0.08$, $c=3.3$ and $d=5.2$. This fitting formula is consistent with the observational data very well, especially at $z\lesssim2$, which is good enough for our estimations here. Then assuming
\be
{\rm SFR}(M,z) = f_*(z)\frac{\Omega_{\rm b}}{\Omega_{\rm M}}\frac{1}{t_{\rm s}}M,
\ee
where $t_{\rm s}=10^8\ \rm yr$ is the typical star formation timescale, and $f_*(z)$ is the normalization factor which can be determined by ${\rm SFRD}(z) = \int dM\frac{dn}{dM}{\rm SFR}(M,z)$. After obtaining SFR$(M,z)$, we can derive $L(M,z)$ for the H$\alpha$, [OII] and [OIII] lines using Eq.~(\ref{eq:SFR_Ha}), (\ref{eq:SFR_OII}) and (\ref{eq:SFR_OIII}) respectively, and then calculate their mean intensities with Eq.~(\ref{eq:I_gal}). We find $\bar{I}_{\rm H\alpha}=31.2\pm9.4\ \rm Jy/sr$, $\bar{I}_{\rm [OII]}=17.5\pm5.0\ \rm Jy/sr$ and $\bar{I}_{\rm [OIII]}=35.0\pm17.0\ \rm Jy/sr$. The errors are derived from the uncertainties in the conversions of the SFR and luminosity given by Eq.~(\ref{eq:SFR_Ha}), (\ref{eq:SFR_OII}) and (\ref{eq:SFR_OIII}).

Another way to obtain the SFR$(M,z)$ is with simulations. Here we use the galaxy catalog from \cite{Guo11}, which is obtained by Millennium II simulation with a volume of $100\ ({\rm Mpc}/h)^3$ and particle mass resolution $\sim 6.9\times10^6\ h^{-1}M_{\sun}$ \citep{Boylan-Kolchin09}. We fit the SFR$(M,z)$ derived from this catalog for $z\lesssim 2$ with
\be 
{\rm SFR}(M,z) = 10^{a+bz}\left( \frac{M}{M_1}\right)^c \left( 1+\frac{M}{M_2} \right)^{d},
\ee
where $a=-9.097$, $b=0.484$, $c=2.7$, $d=-4.0$, $M_1=10^8\ M_{\sun}$ and $M_2=8\times10^{11}\ M_{\sun}$. The simulation results and the best fits of the SFR$(M,z)$ at $z=0.5$, $z=1$ and $z=1.5$ are shown in Figure~\ref{fig:SFR}.

We find that the SFR increases quickly with halo mass but becomes flatter for the large halos with $M\gtrsim 10^{12}\ M_{\sun}$. Also, there is relatively large scattering in the relation at small halo mass with $M\lesssim 10^{10}\ M_{\sun}$, which could provide additional uncertainty in the intensity calculation. The SFRD derived from this SFR$(M,z)$ are $2.2\times10^{-2}$, $3.4\times 10^{-2}$ and $6.6\times10^{-2}\ M_{\sun}\rm yr^{-1} Mpc^{-3}$ at $z\sim0.5$, $z\sim0.9$ and $z\sim1.6$, respectively. For comparison, \cite{Hopkins06} based values are $3.7\times10^{-2}$, $5.9\times10^{-2}$ and $9.7\times10^{-2}\ M_{\sun}\rm yr^{-1} Mpc^{-3}$ at the corresponding redshifts. Thus the SFRDs from the simulation are lower than the SFRDs from the observations. Then we estimate the $L(M,z)$ and obtain $\bar{I}_{\rm H\alpha}=16.8\pm 5.0\ \rm Jy/sr$, $\bar{I}_{\rm [OII]}=9.0\pm2.6\ \rm Jy/sr$ and $\bar{I}_{\rm [OIII]}=16.6\pm8.1\ \rm Jy/sr$ using Eq.~(\ref{eq:SFR_Ha}), (\ref{eq:SFR_OII}), (\ref{eq:SFR_OIII}) and Eq.~(\ref{eq:I_gal}). These values are lower than what we have from the observational SFRD$(z)$, but they are still consistent within 1$\sigma$ error.

\subsection{The mean intensity from the luminosity functions}

A direct way to estimate the mean intensities of the H$\alpha$, [OII] and [OIII] lines is to make use of observed LFs. In Figure~\ref{fig:LF}, we show the observed LFs of H$\alpha$, [OII] and [OIII] lines around $z=0.5$, $z=0.9$ and $z=1.6$ that can contaminate the Ly$\alpha$ emission from $z\sim 7$. The squares, pentagons, circles and triangles are LFs from Ly et al. (2007)\footnote{We note that the H$\alpha$ and [OII] LFs from \cite{Ly07} are well consistent with recent observations with larger survey volumes \cite[e.g.][]{Sobral12,Sobral13}. So we just show the LF data from \cite{Ly07} here.}, Drake et al. (2013) and Gunawardhana et al. (2013), respectively. The LFs data points shown here are dust extinction-corrected, except for the LFs from \cite{Drake13} which are lower than the results of the other two works.

\begin{table}[!t]
\caption{The mean intensities in Jy/sr of the H$\alpha$, [OIII] and [OII] around $z=0.5$, $z=0.9$ and $z=1.6$, derived from both of the LF and SFR methods.}
\vspace{-4mm}
\begin{center}
\begin{tabular}{l | c | c | c  }
\hline\hline
           & $\bar{I}_{\rm H\alpha}(z\sim0.5)$ & $\bar{I}_{\rm [OIII]}(z\sim0.9)$ & $\bar{I}_{\rm [OII]}(z\sim1.6)$\\
\hline 
$\rm SFRD_{obs}$ & $31.2\pm9.4$ & $35.0\pm17.0$ & $17.5\pm5.0$ \\ 
$\rm SFR_{sim}$ & $16.8\pm 5.0$ & $16.6\pm8.1$ & $9.0\pm2.6$ \\
$\rm LF_{L07}$ & $12.2^{+17.4}_{-7.1}$ & $11.8^{+11.1}_{-5.7}$ & $23.5^{+8.3}_{-5.9}$\\
$\rm LF^1_{D13}$ & 8.4 & 12.7 & 24.2 \\
$\rm LF^2_{D13}$ & 23.0 & 6.0 & 50.6 \\
$\rm LF_{G13}$  & 17.6 & $-$ & $-$\\
\hline
\end{tabular}
\end{center}
\vspace{-2mm}
$-$ L07, D13 and G13 denote \cite{Ly07}, \cite{Drake13} and \cite{Gunawardhana13}.\\
\label{tab:Iv}
\end{table}

\begin{figure}[htpb]
\includegraphics[scale = 0.43]{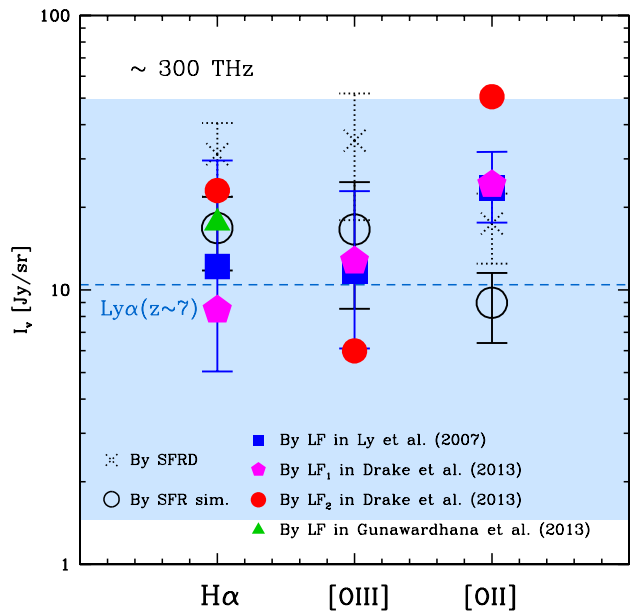}
\caption{\label{fig:Iv} The mean intensity of H$\alpha$, [OIII] and [OII] using both SFR and LF methods around $z=0.5$, $z=0.9$ and $z=1.6$. The dashed line denotes the mean intensity of Ly$\alpha$ at $z\sim7$ with uncertainties (shaded region) derived in the last section. The results from these two methods are basically consistent with each other. The mean intensities of the H$\alpha$, [OII] and [OIII] are generally higher than the central value of the Ly$\alpha$ mean intensity, which can provide considerable contamination on the Ly$\alpha$ emission.}
\end{figure}

The LF is usually fitted by the Schechter function \citep{Schechter76}
\be\label{eq:LF}
\Phi(L)dL = \phi_*\left( \frac{L}{L_*} \right)^{\alpha} {\rm exp}\left(-\frac{L}{L_*}\right)\frac{dL}{L_*},
\ee
where $\phi_*$, $L_*$ and $\alpha$ are free parameters that are obtained by fitting the Schecter function with the data. The LFs from \cite{Ly07} and \cite{Drake13} are fitted by this function\footnote{The H$\alpha$ LF at $0.24<z<0.35$ from \cite{Gunawardhana13} agrees well with the LF at $z=0.4$ from \cite{Ly07} at the faint end, but has flatter slope at the bright end. So it is fitted by an {alternative/different} function with different form given by \cite{Saunders90}.}. Then the mean intensity can be estimated by
\be \label{eq:I_LF}
\bar{I}_{\nu}(z) = \int_{L_{\rm min}}^{L_{\rm max}} dL \frac{dn}{dL} \frac{L}{4\pi D_{\rm L}^2}y(z)D_{\rm A}^2, 
\ee
where $dn/dL=\Phi(L)$ is the luminosity function, and $L_{\rm min}=10^5\ L_{\sun}$ and $L_{\rm max}=10^{12}\ L_{\sun}$ are the lower and upper luminosity limits. We find the result is not changed if choosing smaller $L_{\rm min}$ and larger $L_{\rm max}$ in our calculation. We use the extinction-corrected LFs to calculate the mean intensity for the H$\alpha$, [OII] and [OIII] lines. Note that, in \cite{Drake13}, they just correct the $L_*$ for the dust extinction to get the extinction-corrected LFs, and the other two parameters $\phi_*$ and $\alpha$ are still fixed to the values fitted by the observed LFs.

\begin{figure*}[htpb]
\epsscale{1.9}
\centerline{
\resizebox{!}{!}{\includegraphics[scale=0.29]{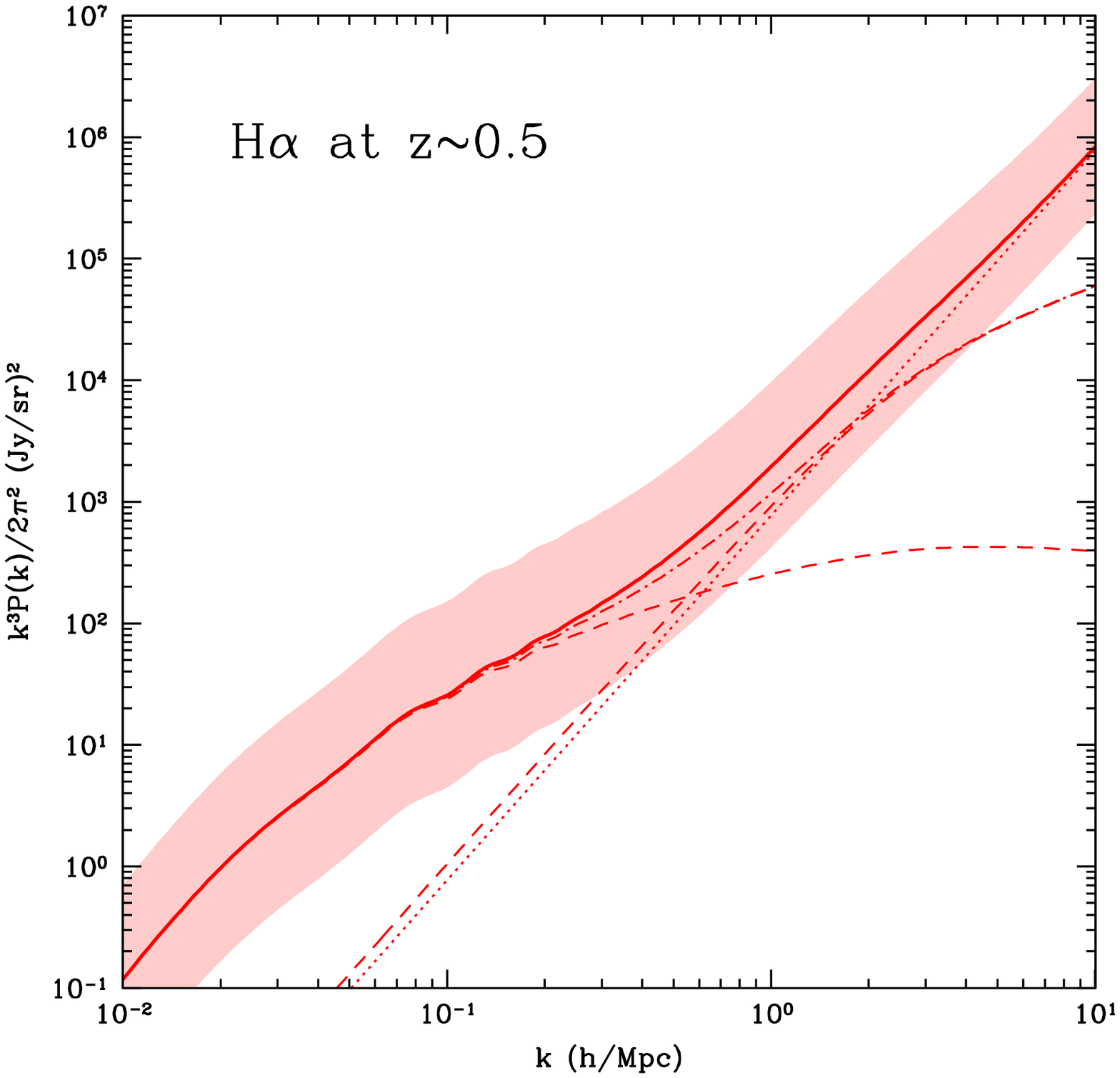}}
\resizebox{!}{!}{\includegraphics[scale=0.29]{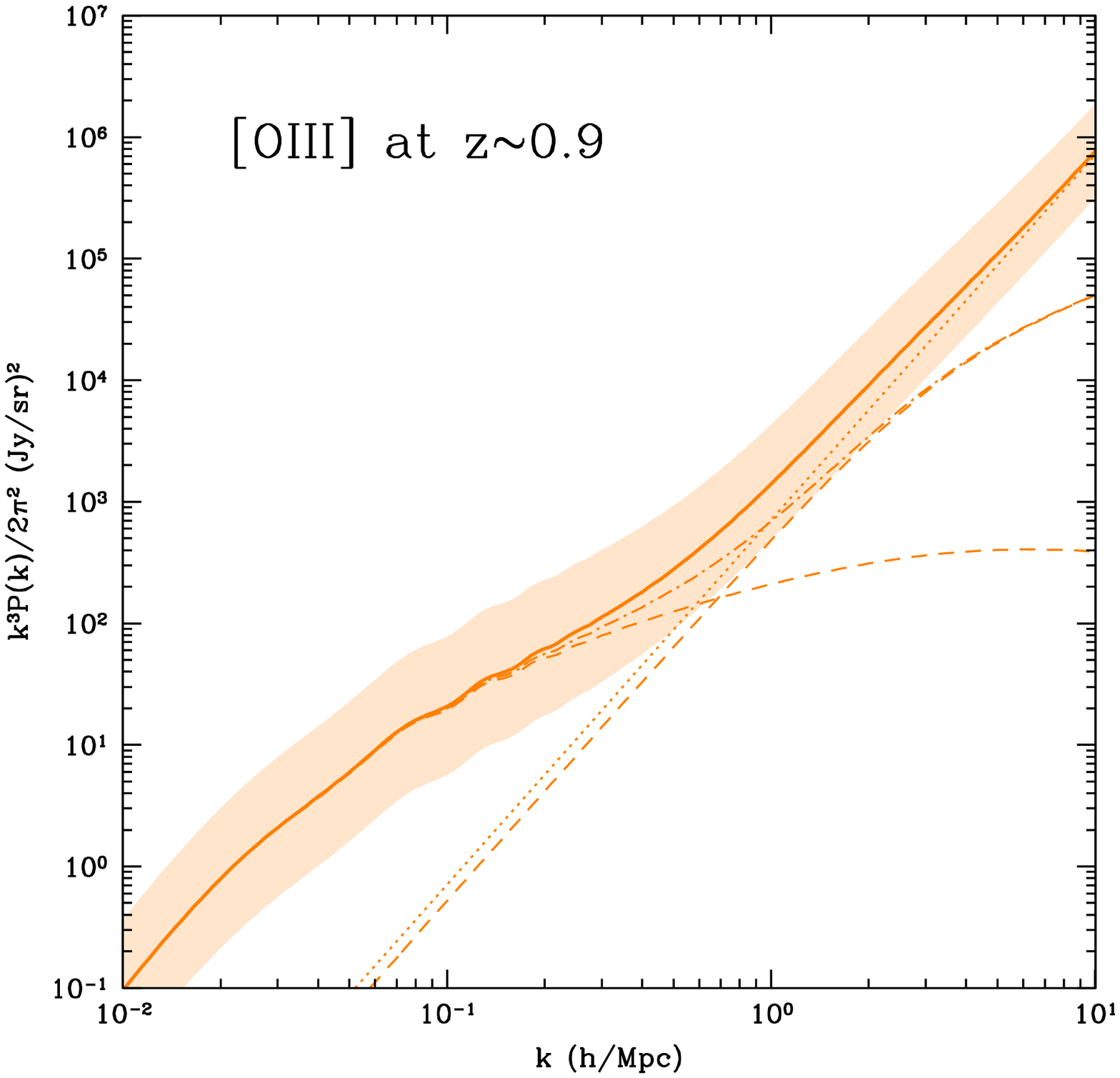}}
\resizebox{!}{!}{\includegraphics[scale=0.29]{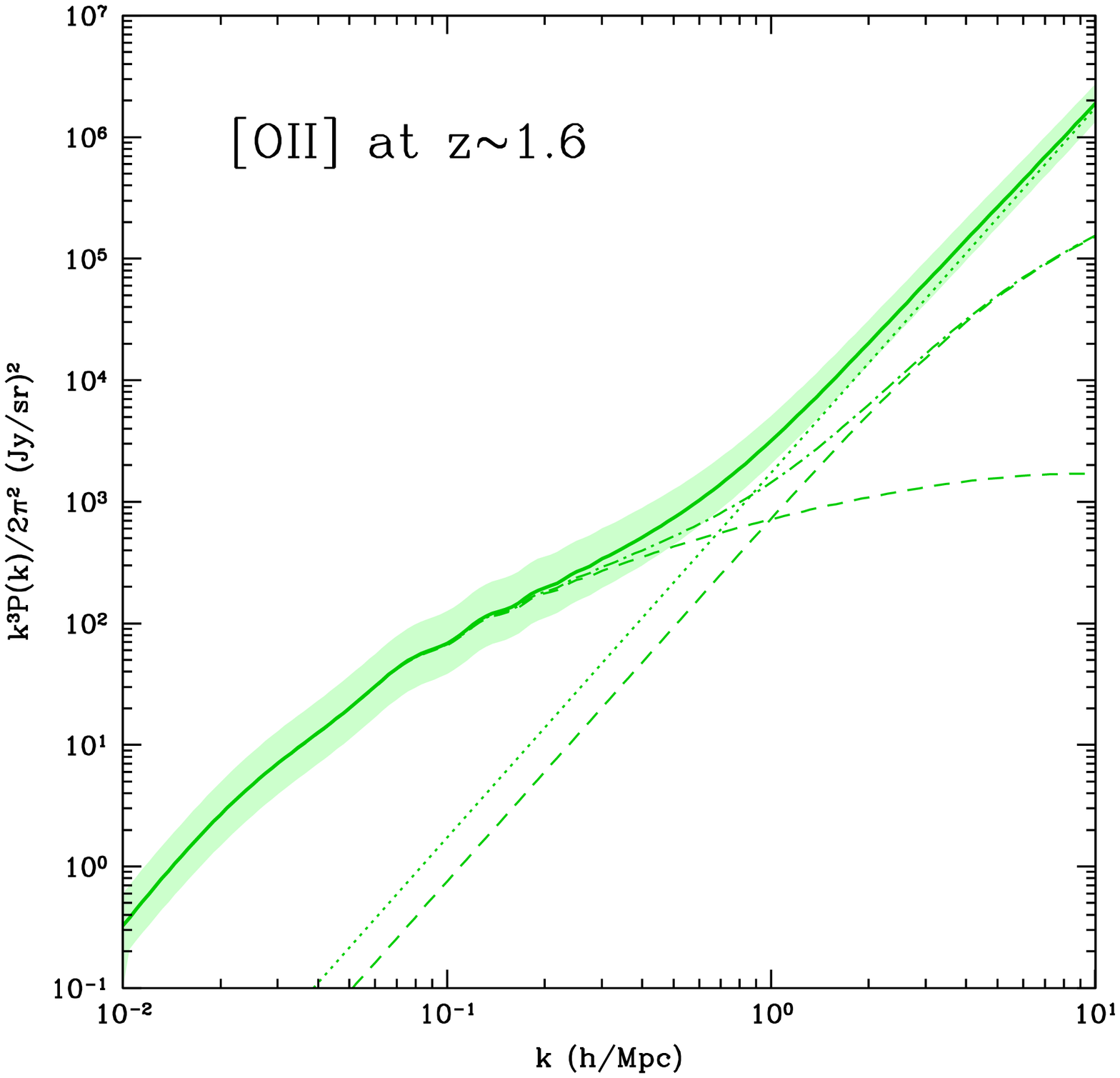}}
}
\epsscale{1.0}
\caption{\label{fig:Pv} The power spectra of the H$\alpha$, [OIII] and [OII] lines at $z\sim0.5$, $z\sim0.9$ and $z\sim1.6$, which can contaminate the Ly$\alpha$ emission at $z\sim7$. The solid, dash-dotted and dotted lines denote the total, clustering and shot-noise power spectrum respectively. The short dashed lines are the 1-halo and 2-halo terms from the halo model \citep{Cooray02}. The uncertainty of the total power spectrum is shown in shaded region, which is estimated by the uncertainty of the mean intensity. Here we adopt the LFs and errors in \cite{Ly07} to estimate the mean intensity and errors for the three emission lines.}
\end{figure*}

Using Eq.~(\ref{eq:I_LF}) and the extinction-corrected LFs, we calculate the mean intensities of the H$\alpha$, [OIII] and [OII] around $z=0.5$, $z=0.9$ and $z=1.6$. We list and plot the mean intensities from both of the LF and SFR methods in Table~\ref{tab:Iv} and Figure~\ref{fig:Iv} for comparison. We also estimate the errors of the LFs from \cite{Ly07} based on the errors of the fitted values of the $\phi_*$, $L_*$ and $\alpha$, and then derive the error for the mean intensity. In Table~\ref{tab:Iv}, the $\rm SFRD_{obs}$ and $\rm SFR_{sim}$ denote the methods of SFRD from observations and SFR from the simulations respectively. The $\rm LF_{L07}$, $\rm LF_{D13}$ and $\rm LF_{G13}$ denote the LF method using the LFs from \cite{Ly07}, \cite{Drake13} and \cite{Gunawardhana13} respectively. The $\rm LF^1_{D13}$ and $\rm LF^2_{D13}$ denote to use the LFs from \cite{Drake13} shown in pink pentagons and red circles in Figure~\ref{fig:LF} respectively. In Figure~\ref{fig:Iv}, the dotted crosses and open circles 
are the values for $\rm SFRD_{obs}$ and $\rm SFR_{sim}$, and the blue squares, pink pentagons, red circles and green triangles are the values for  $\rm LF_{L07}$, $\rm LF^1_{D13}$, $\rm LF^2_{D13}$ and $\rm LF_{G13}$ respectively. For comparison, we also shown the mean intensity of the Ly$\alpha$ at $z\sim7$ (blue dashed line) with uncertainties (shaded region) derived in the last section. 

For the mean intensity of the H$\alpha$, we find the results from the different LF observations are consistent with each other, and they are safely in the $1\sigma$ error of the result from $\rm LF_{L07}$. Comparing the results from the SFR and LF methods, the mean intensity from $\rm SFR_{sim}$ is in a good agreement with that from $\rm LF_{L07}$, $\rm LF^2_{D13}$ and $\rm LF_{G13}$ which give an average H$\alpha$ intensity around $15$ Jy/sr. The $\rm SFRD_{obs}$ and $\rm LF_{D13}^1$ here give a bit higher and lower results, respectively. 

For the [OIII] line, the result of $\rm LF_{L07}$ agrees with the $\rm LF_{D13}^1$ result, and they are also consistent with the result from $\rm SFR_{sim}$ in $1\sigma$ error. This gives an average intensity around $13$ Jy/sr. The $\rm SFRD_{obs}$ gives a higher intensity considering its error. The $\rm LF^2_{D13}$ provides a low [OIII] intensity about 6 Jy/sr, and this can be caused by the poor model fitting of the LF data in \cite{Drake13} (circles in the middle panel of Figure~\ref{fig:LF}). 

For the [OII] emission at $z\sim1.6$, the results have larger discrepancy compared to the H$\alpha$ and [OIII] cases. The $\rm LF_{L07}$ and $\rm LF_{D13}^1$ still give the same intensity which is around 24 Jy/sr, and their results agree with $\rm SFRD_{obs}$ in $1\sigma$. But the $\rm SFR_{sim}$ suggests a low intensity around 9 Jy/sr, while the $\rm LF_{D13}^2$ gives a much higher value of $\sim 50$ Jy/sr. It can be the same reason for the $\rm LF_{D13}^2$ which gives a different result from the others as in the [OIII] case, that there are not enough LF data for the fitting of the LF. In the right panel of Figure~\ref{fig:LF}, we find there are just three data points around $2\times10^{42}\ \rm erg\,s^{-1}$ and no data is observed in the faint end (see red circles). For the $\rm SFR_{sim}$, it actually has large discrepancy between different observations for the SFR-$L_{\rm[OII]}$ relation. For example, \cite{Kewley04} proposes anther relation $${\rm SFR}\,(M_{\sun}{\rm yr^{-1}}) = (6.58\pm 1.65)\times 10^{-42} L_{\rm [OII]},$$ which gives a higher [OII] intensity $\sim 17.4$ Jy/sr using the $\rm SFR_{sim}$ method. This substantially reduces the tension between the $\rm SFR_{sim}$ and the other methods.

Generally, we find that the intensities of the H$\alpha$, [OII] and [OIII] around $z=0.5$, $z=0.9$ and $z=1.6$ are larger than that of the Ly$\alpha$ at $z\sim7$, which provide considerable contamination on the Ly$\alpha$ emission from the EoR. Also, the result derived by the LFs from \cite{Ly07} is well consistent with the other results from both of the LF and SFR method, hence we would adopt it as the foreground line contamination in our following estimation and discussion.

Next, we calculate the anisotropy power spectra of the H$\alpha$, [OII] and [OIII] emissions using their mean intensities. Using Eq.~(\ref{eq:P_Lya}) and (\ref{eq:Pshot_Lya}) and replacing the $\bar{b}_{\rm Ly\alpha}$, $\bar{I}_{\rm Ly\alpha}$ and $L_{\rm Ly\alpha}$ to be the bias, mean intensity and line luminosity of the H$\alpha$, [OII] and [OIII], we obtain their clustering and shot-noise power spectra. Note that we assume the line luminosity is proportional to the halo mass to get the mean bias of these three lines, i.e. replacing $L_{\rm Ly\alpha}$ to be $M$ in Eq.~(\ref{eq:b_Lya}). This approximation is good enough to estimate the mean bias given the uncertainty in the mean intensity. In Figure~\ref{fig:Pv}, we show the power spectrum of the H$\alpha$, [OIII] and [OII] lines at $z=0.5$, $z=0.9$ and $z=1.6$, which contaminate the Ly$\alpha$ emission at $z\sim7$. The total, clustering and shot-noise power spectra are in solid, dash-dotted and dotted lines respectively. The 1-halo and 2-halo terms from the halo model are also shown in short dashed lines \citep{Cooray02}. The uncertainty of the total power spectrum in shaded region is estimated by the uncertainty of the mean intensity. As mentioned, we use the LFs and errors from \cite{Ly07} to compute the power spectrum and the uncertainty for the three emission lines.

\section{The removal of the foreground emission line contamination}

\subsection{The observed power spectrum}
\label{obs_power}
So far we have analyzed the expected intensity from contaminating lines. In practice, experiments (at least first generation ones) will try to make a statistical measurement of the 3-d power spectrum. Therefore, this is the quantity that we should compare to in order to access the level of foreground contamination. Because the signal and foregrounds will be emitted at different redshifts there will be an extra factor multiplying the foreground power spectrum. In order to calculate this, we need to take into account the effect of the observed light cone in the 3-D power spectrum.

The intensity that we measure, $I(\Omega,\nu)$ will correspond to a sum of the
signal we are interested in, $I_s$ (the Ly$\alpha$ emission from redshift $z_s$),
and the foreground emission, $I_f$, from lower redshifts $z_f$, which is
contributing to the same frequency:
\be
I(\Omega,\nu) = I_s(\Omega,\nu) + I_f(\Omega,\nu).
\ee
In the flat sky approximation, displacements in angle $\Delta\theta$ and frequency
$\Delta\nu$ about a central reference point, can be transformed into a position in
3-D comoving space
\ba
&&x_1, x_2 = r(z) \Delta\theta\\\nonumber
&&x_3 = - y(z)\Delta\nu,
\ea
where we are already making a translation along the $x_3$ direction and $x_1, x_2$
are assumed perpendicular to the line of sight while $x_3$ is taken to be parallel to it. 
Note that, besides the flat sky, we are also neglecting cosmic evolution,
e.g. assuming that points along the line of sight are all emitted at the same
redshift $z$.

\begin{figure*}[htpb]
\epsscale{1.9}
\centerline{
\resizebox{!}{!}{\includegraphics[scale=0.37]{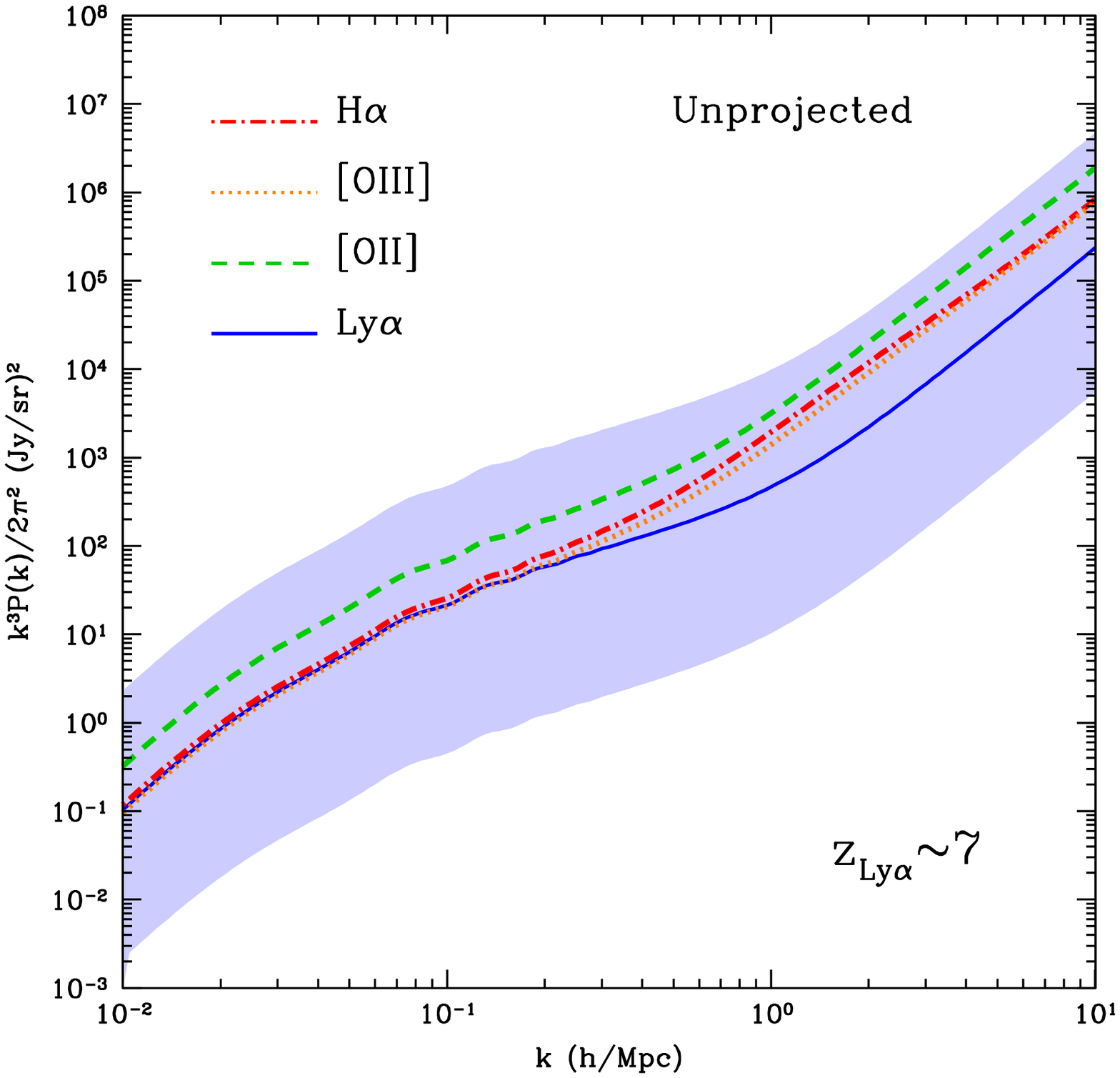}}
\resizebox{!}{!}{\includegraphics[scale=0.37]{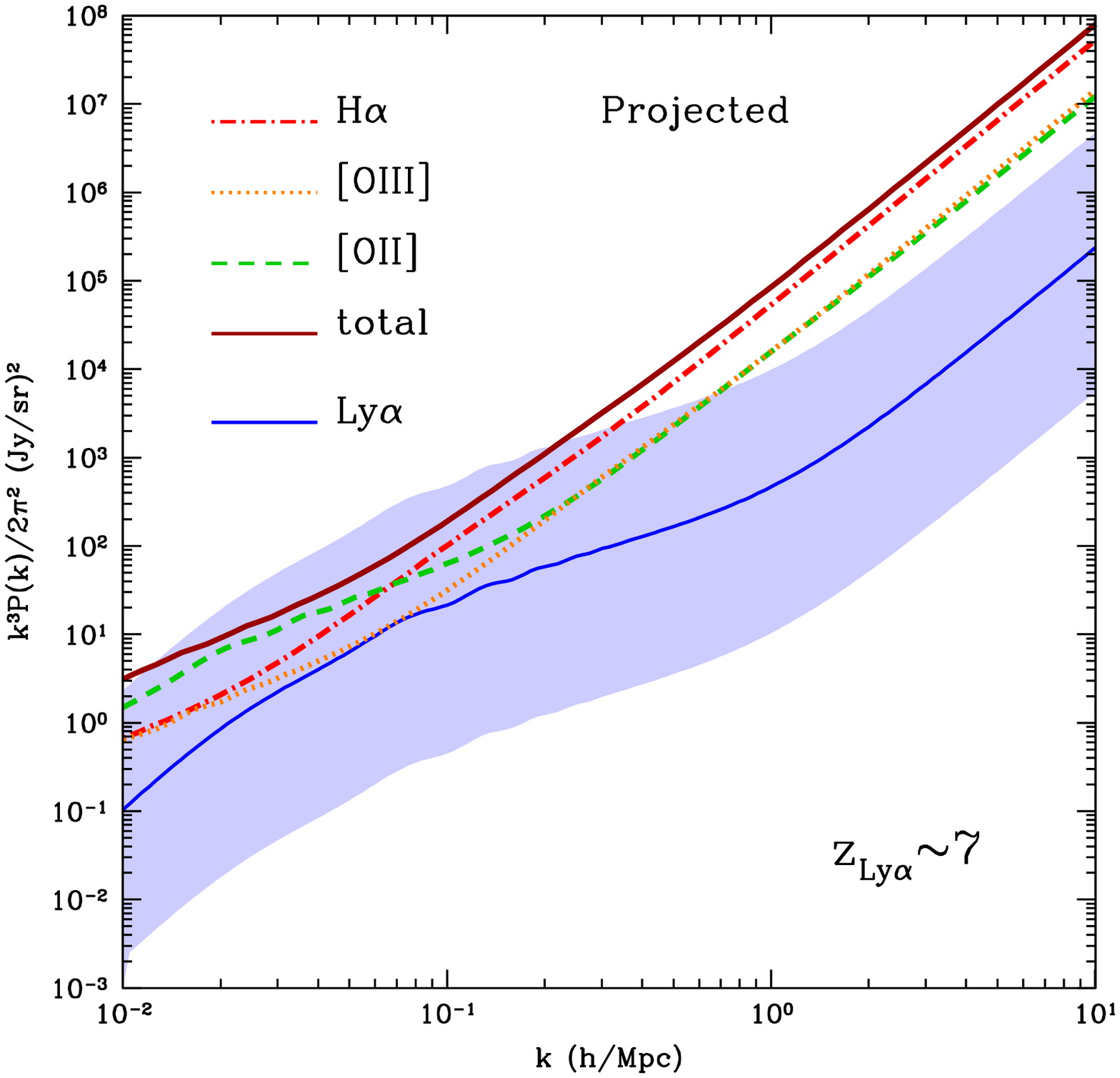}}
}
\epsscale{1.0}
\caption{\label{fig:Pcon} $Left:$ the comparison of the 3-D total power spectrum of the Ly$\alpha$ at $z\sim7$ to that of the H$\alpha$ at $z=0.5$, [OIII] at $z=0.9$ and [OII] at $z=1.6$. The uncertainty of the Ly$\alpha$ power spectrum is shown in shaded region. $Right:$ the same as the left panel, but considering the projection effect on the power spectra of the H$\alpha$, [OII] and [OIII] lines. The total projected power spectrum is shown in dark red line.}
\end{figure*}

We then write the observed signal as
\ba
I(\Omega,\nu) &=& \bar{I}_s(z_s)[1+b_s(z_s)\delta(z_s,{\mathbf x}_s)] \nonumber \\
              &+&\bar{I}_f(z_f)[1+b_f(z_f)\delta(z_f,{\mathbf x}_f)]
\ea
where $\delta()$ is the dark matter perturbation and ${\bf x}_s$, ${\bf x}_f$ is the
3-D position of the signal and foreground emission respectively.
If we Fourier transform the above signal with respect to ${\bf x}_s$, the
corresponding foreground Fourier mode will be offset with respect to the true one.
The observed power spectrum, $P_{\rm obs}$ will then be:
\ba \label{eq:P_obs}
P_{\rm obs}({k}_{\perp}, k_{\parallel}) &=& \bar{I}^2_s(z_s) b^2_s(z_s) P(z_s,k_s) \\ \nonumber
                                &+&\bar{I}^2_f(z_f) b^2_f(z_f) \left(\frac{r_s}{r_f}\right)^2 \left(\frac{y_s}{y_f}\right)P(z_f,k_f),
\ea
where $|\vec{k}_s|=\sqrt{k^2_{\perp}+k^2_{\parallel}}$ is the 3-D $k$ at the redshift of the signal, and $|\vec{k}_f|=\sqrt{(r_s/r_f)^2 k^2_{\perp}+(y_s/y_f)^2 k^2_{\parallel}}$ is the 3-D $k$ at the redshift of the foreground line. The factor $(r_s/r_f)^2(y_s/y_f)$ comes from the distortion of the volume element when Fourier transforming the foreground correlation function to the power spectrum \citep{Visbal10}. This process can be considered to project the foreground power spectrum to the high redshift where the intensity signal comes from. We then see that, even when looking at the spherical averaged power spectrum, e.g. $P_{\rm obs}(k=\sqrt{{k}_{\perp}^2 + k_{\parallel}^2})$, the foreground contribution to the power spectrum will be boosted by this projection. Moreover, there is an anisotropy in the $\mathbf k$ space which provides a potential method to distinguish the foregrounds from the signal. We discuss the details of this effect in the Appendix.

In Figure~\ref{fig:Pcon}, we compare the 3-D power spectrum of H$\alpha$, [OII] and [OIII] with and without the projection. In the left panel of Figure~\ref{fig:Pcon}, we show the total power spectrum of the H$\alpha$ at $z=0.5$, [OIII] at $z=0.9$ and [OII] at $z=1.6$ without projection. The $P_{\rm Ly\alpha}^{\rm tot}$ and its uncertainty are also shown in blue line and shaded region for comparison. In the right panel of Figure~\ref{fig:Pcon}, we show the 3-D projected power spectra of the H$\alpha$, [OII] and [OIII] lines to the redshift of the Ly$\alpha$ emission ($z\sim7$). To estimate foreground 3-D projected power spectrum from Eq.~(\ref{eq:P_obs}), we simply assume $k_1=k_2=k_{\parallel}$, where ${\mathbf k}=k_1\hat{\mathbf i}+k_2\hat{\mathbf j}+k_{\parallel}\hat{\mathbf k}$ and $k_{\perp}=\sqrt{k_1^2+k_2^2}$. We also ignore the redshift distortion here. We find the projected power spectra of the three lines become a bit higher than the case where we igore the projection of the foregrounds to the background cosmological signal of intensity during the EoR.

\subsection{Foregrounds masking}

To remove the contamination from these foreground emission lines, we need to mask the bright sources at low redshifts. However, we cannot identify the individual sources and their redshifts in the intensity mapping, and all the signals that lie in the same survey pixel, which is defined by the spectral and angular resolutions, will be mixed together and observed as only one signal.

\begin{figure}[htpb]
\includegraphics[scale = 0.4]{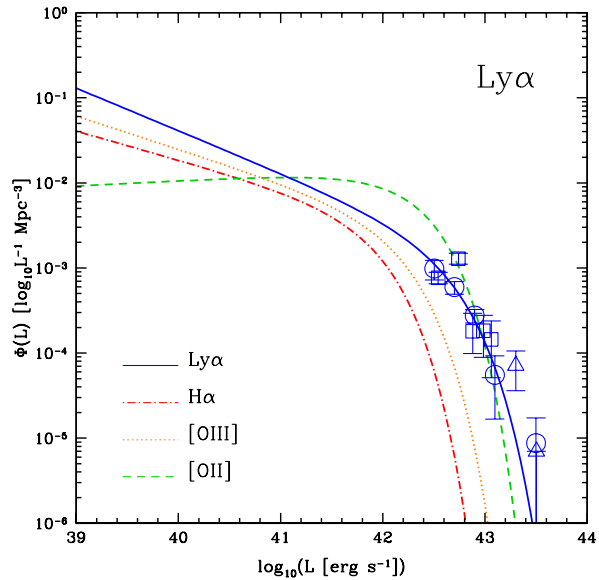}
\caption{\label{fig:LF_Lya} The LF of the Ly$\alpha$ at $z\sim7$ (actually around z=6.6) from the observations of the LAEs. The circles, squares and triangles are the LF data points from \cite{Ouchi10}, \cite{Kashikawa06} and \cite{Hu05}, respectively. The blue solid curve is the best fit of the LF data using the fitting results given in \cite{Ouchi10}. For comparison, the LFs of the H$\alpha$, [OIII] and [OII] around $z=0.5$, $z=0.9$ and $z=1.6$ from \cite{Ly07} are also shown.}
\end{figure}

During the EoR, the galaxies are averagely smaller and fainter than the present galaxies, so their LFs should be relatively higher at the faint end than the galaxies at the low redshifts. In Figure~\ref{fig:LF_Lya}, we show the Ly$\alpha$ LF at $z\sim7$ from the observations of the Ly$\alpha$ emitters (LAEs) \citep{Ouchi10,Kashikawa06,Hu05}\footnote{Using Eq.~(\ref{eq:I_LF}) and the Ly$\alpha$ LF (blue solid curve in Figure~\ref{fig:LF_Lya}), we note that the Ly$\alpha$ mean intensity is just $\sim1$ Jy/sr around $z=7$, which is lower than the value of $\sim9$ Jy/sr from the galaxy we get in Section 2 (see also Pullen et al. 2013). This can be caused by the absence of the Ly$\alpha$ LF data at the faint end, since only very bright individual sources can be detected at this high redshift. This also means that there could be more faint LAEs around $z=7$, and the real LF could be higher at the faint end.}. The best fit of the LF is shown in blue solid line using the Schechter function (see Eq.~(\ref{eq:LF})) with fitting values given by \cite{Ouchi10}. The LFs of the H$\alpha$, [OIII] and [OII] lines are also shown for comparison. We can see that, the LFs of the Ly$\alpha$ is higher than the others at the faint end, but basically lower than the [OII] LF at the bright end. Therefore, we can mask the whole bright pixels whose fluxes are above some threshold in the intensity mapping to reduce the contamination of low-redshift line emission, and then the remainder pixels should be dominated by the Ly$\alpha$ emission at high redshifts.

\begin{figure}[htpb]
\includegraphics[scale = 0.4]{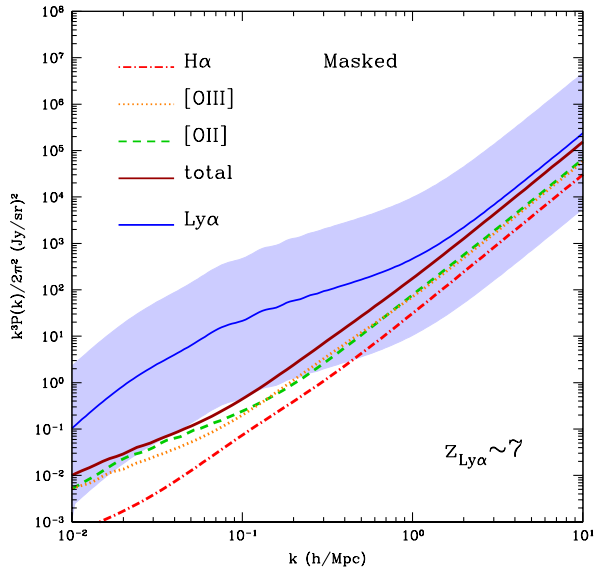}
\caption{\label{fig:Pcon_mask} The power spectra of the H$\alpha$, [OII] and [OIII] with masking and projection. The deep red line denotes the total foreground power spectrum after the masking. We apply a flux cut at $1.4\times10^{-20}$ W/m$^2$ here, which could make the total foreground power spectrum $\sim$100 times smaller than the Ly$\alpha$ around $k=0.1$ $h$ Mpc$^{-1}$.}
\end{figure}

In Figure~\ref{fig:Pcon_mask}, the power spectra of the H$\alpha$, [OII] and [OIII] lines with masking and projection are shown. Here we try to make the total power spectrum of ${\rm H\alpha}$, ${\rm [OII]}$ and ${\rm [OIII]}$ smaller by a factor of $\sim100$ around $k=0.1$ $h$ Mpc$^{-1}$ where the shot-noise is small. We find we need to mask sources with fluxes greater than $1.4\times10^{-20}$ W/m$^2$. The corresponding line luminosity are $L_{\rm H\alpha}\simeq 3.5\times10^6\ L_{\sun}$, $L_{\rm [OII]}\simeq 6.3\times10^7\ L_{\sun}$ and $L_{\rm [OIII]}\simeq 1.5\times10^7\ L_{\sun}$ respectively. Note that these luminosity cuts are close to the lower luminosity limit ($L_{\rm min}=10^5\ L_{\sun}$) we take in Eq. (\ref{eq:I_LF}), hence our masking results are dependent on the faint-end slopes of the LFs which are not well constrained by observational data.

\begin{figure}[htpb]
\includegraphics[scale = 0.4]{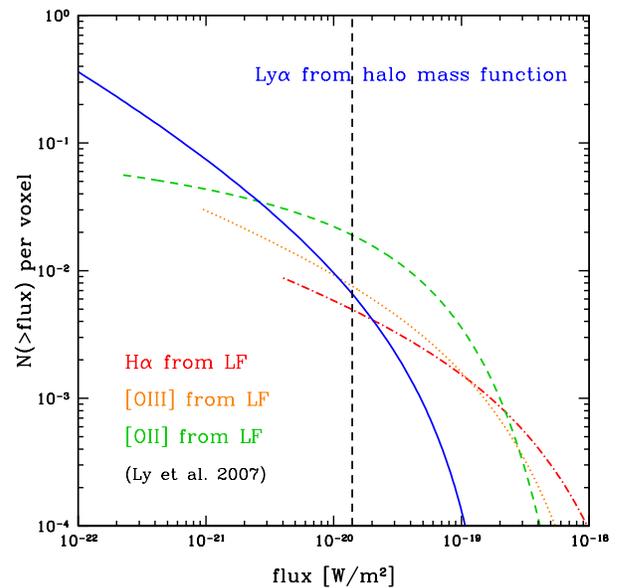}
\caption{\label{fig:N_flux} The number of sources per volume pixel whose fluxes are greater than the x-axis flux value. The Ly$\alpha$ curve is derived from the halo mass function and the calculation in Section 2, and the curves of H$\alpha$, [OII] and [OIII] are computed by the LFs in \cite{Ly07}. The vertical dashed line denotes the flux cut at $1.4\times10^{-20}$ W/m$^2$. We find this corresponds to remove 3\% of the total pixels.}
\end{figure}

In Figure~\ref{fig:N_flux}, we show the number of sources whose flux is greater than some value in a survey volume pixel. Here we assume a survey with 6$''$$\times$6$''$ beam size and $190-330$ THz frequency range with R=40 frequency resolution. As can be seen, the H$\alpha$ sources are dominant at the bright end, but are subdominant in the survey since all of these sources are from low redshifts ($z\sim0.5$) where the survey volume is small. On the other hand, [OII] sources are the main contamination, and are brighter than $1.4\times10^{-20}$ W/m$^2$ in approximately 2\% of the pixels. In total, we need to mask about 3\% survey pixels. Here we should notice the large uncertainty of the Ly$\alpha$ emission. The number of the Ly$\alpha$ sources can be lower by a factor of $\sim10$ with the uncertainty, which would lead to a larger masking percentage.

There is a second method that can be used to eliminate the foreground line contamination, which is making use of the cross-correlation between different emission lines at the same redshift. This method is discussed in many previous works and can be used either for the signal (e.g. Visbal \& Loeb 2010; Gong et al. 2012; Silva et al. 2013) or the foreground contamination (e.g. Pullen et al. 2013). The idea is that the lines emitted at the same redshift should trace the same underlying matter distribution and hence lead to large cross-correlation on the power spectrum. On the other hand, the emissions at different redshifts, which are far away from each other, would not provide considerable cross power spectrum. Therefore, we can derive the auto power spectrum of one line if we know both the cross power spectrum and the auto power spectrum of the other line. For instance, we can cross-correlate the Ly$\alpha$ line with 21-cm at $z\sim7$, and then the foreground contamination would be reduced and we can estimate the $P_{\rm Ly\alpha}$ from the $P_{\rm Ly\alpha \times 21cm}$ and $P_{\rm 21cm}$ \citep{Silva13}. Or we can estimate the auto power spectra of the foreground emission lines by cross-correlating with the 21-cm intensity mapping surveys at $z\sim1$, e.g. the GBT\footnote{https://science.nrao.edu/facilities/gbt/}, CHIME\footnote{http://chime.phas.ubc.ca/} and Tianlai projects \citep{Chen12}. However, although we can detect the cross power spectrum, it is still hard to measure the auto power spectrum for the 21-cm or the other lines due to their own foregrounds removal. So this method is indirect and cannot substitute the masking method discussed above. Even if we can only obtain the cross power spectrum, it could still be a good guide and a secondary check for the intensity mapping experiments.

\section{Summary}

We estimate the Ly$\alpha$ mean intensity and power spectrum during the EoR, and explore the foreground contamination from the low-redshift emission lines. We consider the Ly$\alpha$ emission from both of galaxies and IGM for the recombination and collisional emission processes. We find the Ly$\alpha$ emission of galaxies is dominant over the IGM at $z\sim7$ with a total mean intensity $\bar{I}_{\rm Ly\alpha}\sim 10$ Jy/sr (about 9 Jy/sr from galaxies and 1 Jy/sr from the IGM). We also evaluate the uncertainty of the mean intensity according to the uncertainties of $f_{\rm esc}^{\rm ion}$, $f_{\rm Ly\alpha}$, SFR and the clumping factor. With the help of halo model, we also calculate the Ly$\alpha$ clustering, shot-noise power spectrum with the uncertainty given by the mean intensity.

Next we investigate the foreground contamination by low-redshift emission lines. We find H$\alpha$ at $6563\ \rm \AA$, [OIII] at $5007\ \rm \AA$ and [OII] at $3727\ \rm \AA$ can be the strongest contamination on Ly$\alpha$ emission during the EoR. We estimate the mean intensity of the H$\alpha$, [OIII] and [OII] lines at $z\sim0.5$, $z\sim0.9$ and $z\sim1.6$ respectively, which are the redshifts that can contaminate the Ly$\alpha$ emission at $z\sim7$. We use two methods to do the estimation, i.e. the SFR and LF methods. In the SFR method, both of the SFRD$(z)$ from the observations and the SFR$(M,z)$ derived from the simulations are used to compute the intensity. In the LF method, we adopt the LFs of H$\alpha$, [OIII] and [OII] around $z=0.5$, $z=0.9$ and $z=1.6$ respectively from different observations. We find the results from both methods are basically consistent with each other, especially for the H$\alpha$ and [OIII] lines. The mean intensity of the three lines are $\bar{I}_{\rm H\alpha}\sim 15$ Jy/sr, $\bar{I}_{\rm [OIII]}\sim 13$ Jy/sr and $\bar{I}_{\rm [OII]}\sim 24$ Jy/sr, which are larger than the $\bar{I}_{\rm Ly\alpha}\sim 10$ Jy/sr. The results from the LFs in \citep{Ly07} is in good agreements with the others, and we adopt their LFs and the errors to calculate the power spectrum and uncertainty for the three foreground lines.

At last, we discuss the methods to remove the foreground contamination due to low-redshift emission lines. We first compare the power spectrum of the Ly$\alpha$ at $z\sim7$ to the H$\alpha$, [OIII] and [OII] power spectrum at low redshifts, and consider the projection effect in the real survey. The power spectrum of the foreground lines become larger after the projection. We then propose to mask the whole bright pixels with foreground emission above some flux threshold to reduce the contamination of the foreground lines in the intensity mapping. We find the contamination can be neglected when the flux cut is $1.4\times 10^{-20}\ \rm W/m^2$. The cross-correlation method is helpful to reduce or estimate the contamination, but it is indirect and can not substitute the masking method in the survey of the intensity mapping.

\begin{acknowledgments}
This work was supported by NSF CAREER AST-0645427 and AST-1313319.
MBS and MGS acknowledge support by FCT-Portugal under grant PTDC/FIS/100170/2008.
MBS acknowledges support by FCT-Portugal under grant SFRH/BD/51373/2011.
\end{acknowledgments}

\appendix

\begin{figure*}[htpb]
\epsscale{1.9}
\centerline{
\resizebox{!}{!}{\includegraphics[scale=0.3]{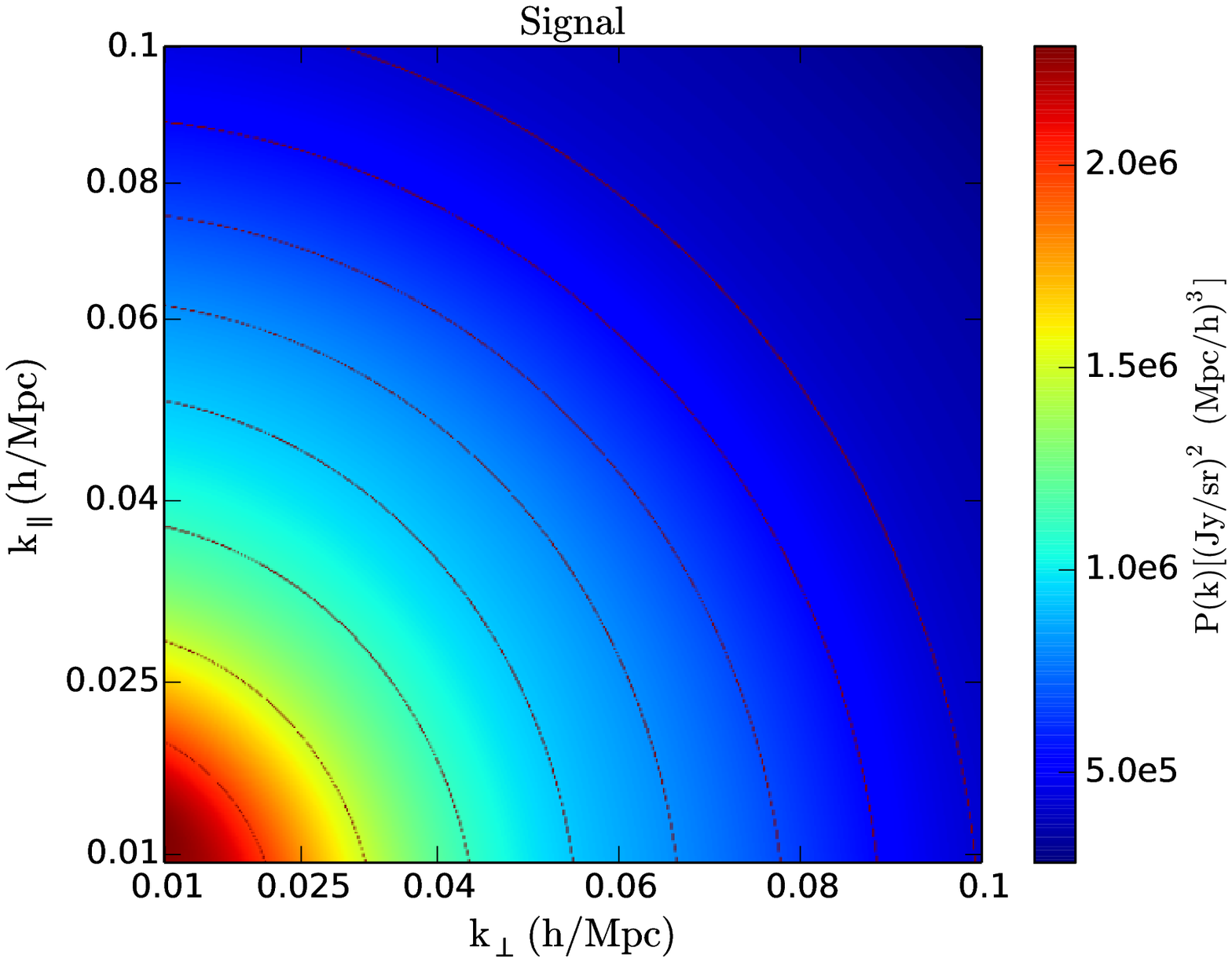}}
\resizebox{!}{!}{\includegraphics[scale=0.3]{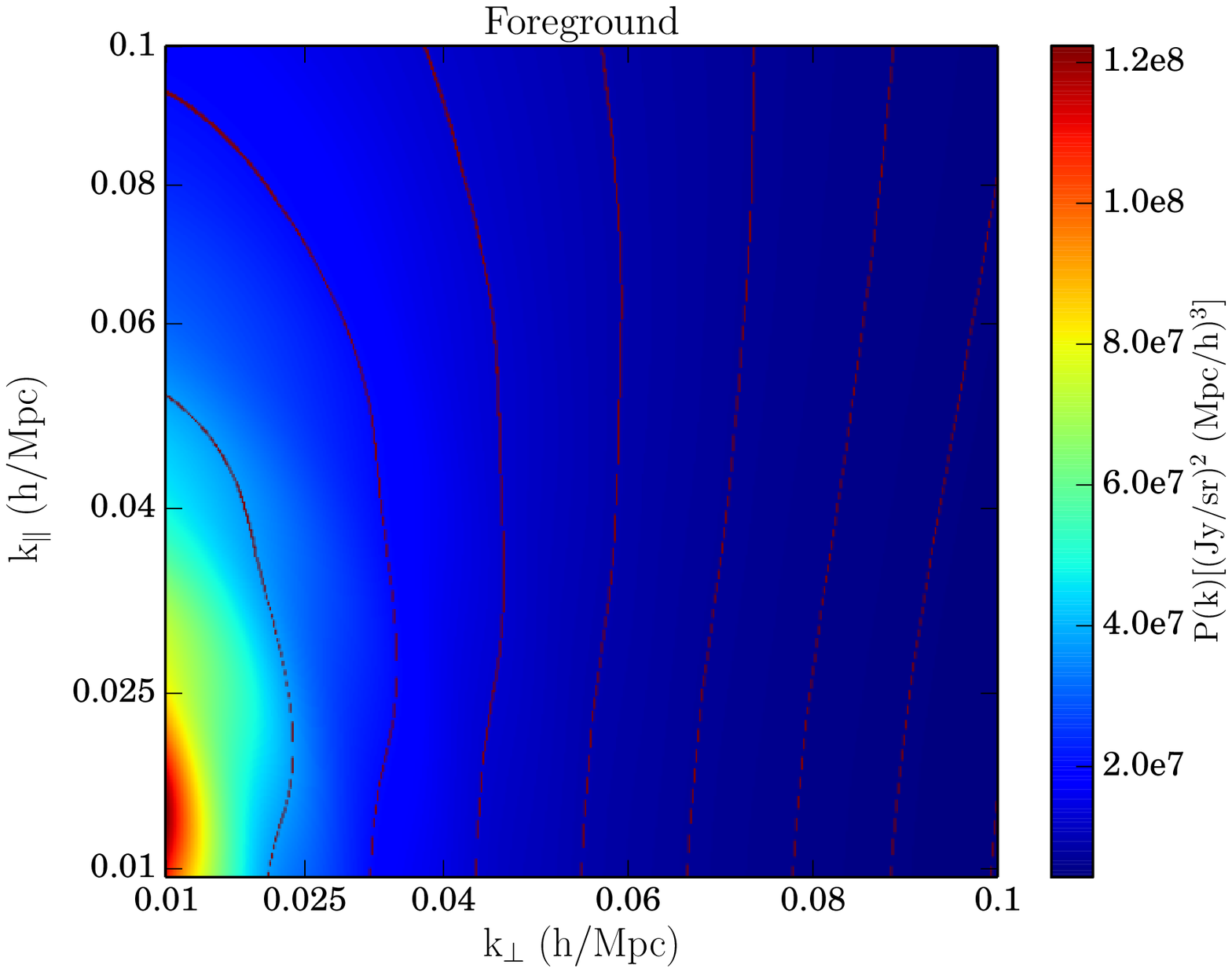}}
\resizebox{!}{!}{\includegraphics[scale=0.3]{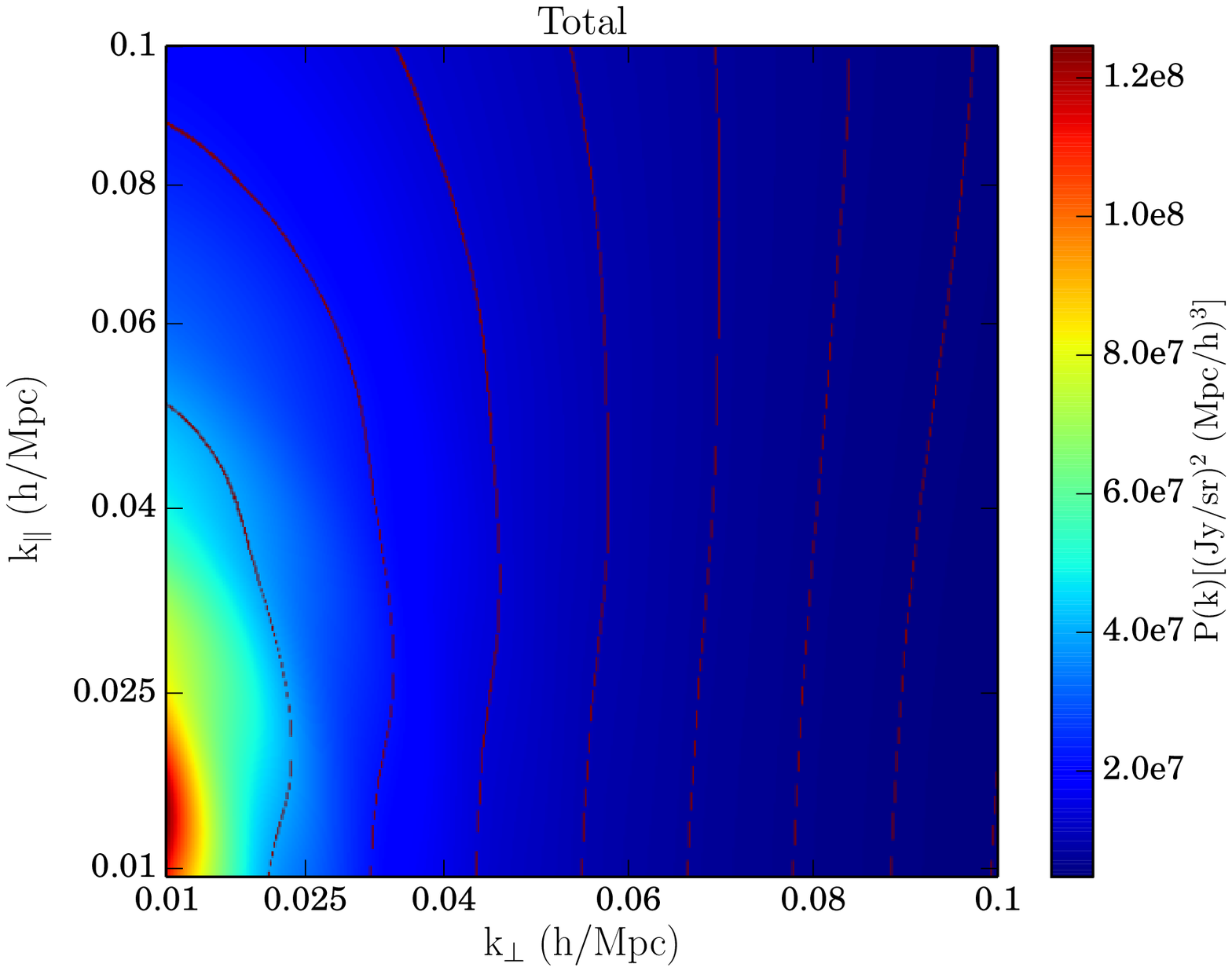}}
}
\epsscale{1.0}
\caption{\label{fig:P_per_par} The 2-D anisotropic power spectra shown by perpendicular and parallel Fourier modes for the signal (Ly$\alpha$), foreground (only H$\alpha$ here) and total observed signal+foreground. We find the 2-D signal power spectrum is almost symmetric for $k_{\perp}$ and $k_{\parallel}$, and the redshift distortion effect is relatively small. For the foreground power spectrum at low redshift, because the redshift distortion effect is strong and the shift factor on the foreground $k_{\perp}$ and $k_{\parallel}$ are different, the shape of the power spectrum is irregular. This effect provides a way to distinguish the signal and the foreground in principle. The total power spectrum is similar with the foreground power spectrum, since the amplitude of the signal power spectrum is much lower than the foreground.}
\end{figure*}

In Eq. (\ref{eq:P_obs}) of Section \ref{obs_power}, we see that the observed foreground power spectrum is actually anisotropic, i.e., not simply a function of $k=\sqrt{{k}_{\perp}^2 + k_{\parallel}^2}$. This can in principle be used in the foreground cleaning process. In particular, after the intensity cut, we can check if there is still any strong contamination by looking at the anisotropy of the total power spectrum.
In order to check the strength of these anisotropies caused by the projection effect, we calculate the 2-D anisotropic power spectra for the signal and foregrounds. Here we also take into account of the linear redshift-space distortions. In that case, the bias should be replaced by
\ba
&&b_s(z_s) \rightarrow
b_s(z_s)+\left(\frac{k_{\parallel}}{\sqrt{k_{\perp}^2+k_{\parallel}^2}}\right)^2\frac{1}{H(z_s)}\frac{\dot{D}(z_s)}{D(z_s)}\nonumber\\
&&b_f(z_f) \rightarrow b_f(z_f)+
\left(\frac{(y_s/y_f)k_{\parallel}}{\sqrt{(r_s/r_f)^2k_{\perp}^2+(y_s/y_f)^2k_{\parallel}^2}}\right)^2\frac{1}{H(z_f)}\frac{\dot{D}(z_f)}{D(z_f)}.\nonumber
\ea
Here, $D(z)$ is the growth factor. The normal expression of the second term of the bias is $f\mu^2$. Here $f=d\,{\rm ln}D/d\,{\rm ln}a$ where $a$ is the scale factor, and $\mu={\rm cos}\theta$ where $\theta$ is the angle between the line of sight and the wave-vector ${\mathbf k}$. We also notice that the so-called non-gravitational effects can introduce strong redshift-space distortion effect \citep{Zheng11,Wyithe11}. According to full Ly$\alpha$ radiative transfer calculations, the Ly$\alpha$ emission is dependent on environment (gas density and velocity) around LAEs. The observed LAE clustering features can be changed by this effect especially at high redshifts. However, this effect relies on ``missing'' Ly$\alpha$ photons scattering in relatively close proximity to Ly$\alpha$ emission galaxies, which could be recovered by intensity mapping. Therefore we ignore this effect in our discussion.

In Figure~\ref{fig:P_per_par}, we show the 2-D anisotropic power spectrum decomposed into $k_{\perp}$ and $k_{\parallel}$ for the signal, foreground and the total observations. We just show the H$\alpha$ foreground here, since it has the lowest redshift among the other two foreground lines and has the largest effect of the power spectrum projection. We find the shape of the signal power spectrum is quite symmetric and the redshift distortion effect is relatively small, since the signal comes from high redshifts. However, for the foreground, the shape of the spectrum is irregular due to the redshift distortion and the different factors on the $k_{\perp}$ and $k_{\parallel}$ (i.e. $r_s/r_f$ on $k_{\perp}$ and $y_s/y_f$ on $k_{\parallel}$) in the projection. This effect provides a potential method to distinguish the signal from the foreground, and could be helpful to remove the foreground in future high sensitive experiments.

\end{document}